\documentclass{article}
\usepackage{PRIMEarxiv}

\usepackage{multirow}
\usepackage[table]{xcolor}
\usepackage{enumitem}

\usepackage[utf8]{inputenc} 
\usepackage[T1]{fontenc}    
\usepackage{hyperref}       
\usepackage{url}            
\usepackage{booktabs}       
\usepackage{amsfonts}       
\usepackage{nicefrac}       
\usepackage{microtype}      
\usepackage{lipsum}
\usepackage{fancyhdr}       
\usepackage{graphicx}       
\graphicspath{{media/}}     
\usepackage{authblk}
\usepackage{subcaption}

\usepackage{multirow}
\usepackage{pdfpages}
\usepackage{natbib}
\usepackage{amsmath}

\title{The Significance of User Characteristics for Reposting Prediction on X: A Comparative Analysis Under Distribution Shift}

\author[1]{\textbf{Ziming Xu}}
\author[1]{\textbf{Shi Zhou}}
\author[1]{\textbf{Vasileios Lampos}} 
\author[1,2]{\textbf{Ingemar J. Cox}}

\affil[1]{Centre for Artificial Intelligence, Department of Computer Science, University College London, UK}
\affil[2]{Department of Computer Science, University of Copenhagen, Denmark}

\begin{document}

\maketitle

\begin{abstract}

Understanding information diffusion on X (formerly Twitter) requires accurate modelling of reposting behaviour. Most existing work predicts reposting under in‑distribution settings, where training and test data cover the same topics. This paper addresses a more realistic and challenging scenario: out‑of‑distribution prediction, i.e., forecasting reposting behaviour for new, previously unseen topics. We formulate the task at the individual level — predicting whether a specific user will repost a given post — and systematically compare the predictive power of post‑related features, user‑related features, and their combination across four representative models: Decision Tree, Multi‑Layer Perceptron, BERT, and Qwen. Our experiments show that while post‑related features perform well in‑distribution, their performance declines drastically for unseen topics, with F1 scores falling to approximately 0.12. In contrast, user‑related features — including user profiles, social relations, and historical behaviour — deliver strong and transferable performance, raising the F1 score to over 0.70. These results demonstrate that reposting decisions are largely content‑agnostic: they are driven more by stable user characteristics than by the specific content of a post. Our findings highlight the value of user modelling for building robust prediction systems and provide new insights into the mechanisms that enable information to spread across different topics.
\end{abstract}

\keywords{Online Social Media, Social Content Sharing, Repost/Retweet, Out-of-Distribution Generalisation, Prediction, Machine Learning, X (Twitter)}

\section{Introduction}

Understanding and predicting user reposting behaviour on social networks is an area of interest across disciplines such as computer science, social science, political science, and marketing. This interest stems from the influence that individual behaviours have on information diffusion~\cite{jerez2026human,tang2025msa,nettasinghe2025group}, 
including information exposure and amplification in marketing~\cite{tu2022viral,gu2025online} and political campaigns~\cite{engel2025social,cinus2025exposing}, polarisation and echo chambers~\cite{impicciche2025comparing,nettasinghe2025group}, and how misinformation and rumours spread~\cite{ashkinaze2024dynamics,sakib2025opposites,evangelatos2025modeling}.
On social media platforms, reposting is one of the main observable actions through which users propagate information to others.
Modelling reposting behaviour is therefore an important step towards understanding how information spreads through social networks.

Prior studies have commonly formulated the prediction of reposting as a supervised binary classification task~\cite{xu2012analyzing,luo2013will,yang2017propagator,zhang2013social,feng2013retweet,hoang2016microblogging}. The objective is to train a binary classifier to estimate the likelihood of a recipient reposting a post from a sender. In this formulation, the dataset is typically divided into training and test sets through random assignment, ensuring that both sets are drawn from the same underlying data distribution for fair and stable evaluation. Such in-distribution evaluation is useful for measuring predictive performance when future cases resemble past cases.

On social media platforms, however, new topics appear regularly, e.g., those related to breaking news. Thus, in addition to the typical in-distribution classification, out-of-distribution prediction is also needed, where a model is trained on existing data related to previous topics and is then used to predict a user's reaction to a post related to a new, previously unseen topic.
Under this setting, models cannot rely only on topic-specific vocabulary, entities, or semantic contexts observed during training.
They must instead use signals that transfer across topics, such as properties of the sender, the recipient, their social relation, and their historical behaviour.
This observation motivates our central question: when predicting reposting for unseen topics, do post-related features, user-related features, or their combination provide the most transferable predictive signal?

In this paper, we first review previous work on predicting repostings on X (formerly Twitter).
Existing algorithms predominantly focused on the textual data, i.e., the text content of posts. 
We argue that features of the sender and the recipient are also highly relevant to the reposting prediction and remain underutilised. 
We study two types of information: (i)~the post text, and (ii)~user activity and profile attributes.
We derive features from these data that capture both post-related information and user-related characteristics.
We consider four classification models: a decision tree (DT), a multi-layer perceptron (MLP), BERT~\cite{devlin2019bert}, and Qwen~\cite{yang2025qwen3technicalreport}.
Using data collected from X, we conduct experiments under both in-distribution and out-of-distribution evaluations.   

Our results show that while the prediction models, using only post-related features, perform well for in-distribution prediction, they perform much worse for out-of-distribution prediction. 
When post-related features are supplemented or replaced with user-related features, the out-of-distribution performance greatly improves, with the F1 score increasing from approximately 0.12 to over 0.70.
 
Our work highlights the importance of user-related features for reposting prediction---features that are currently underutilised~\cite{zhang2016retweet,ma2019hot,jiang2023retweet,guo2025somermultiviewuserrepresentation}.
Interestingly, our experimental results suggest that reposting decisions are largely content‑agnostic: they are driven more by stable user characteristics than by the specific content of a post. 
This work provides novel insights into the mechanisms that drive information diffusion on social networks.
 
\begin{figure}[!t]
\centering
\includegraphics[width=\textwidth]{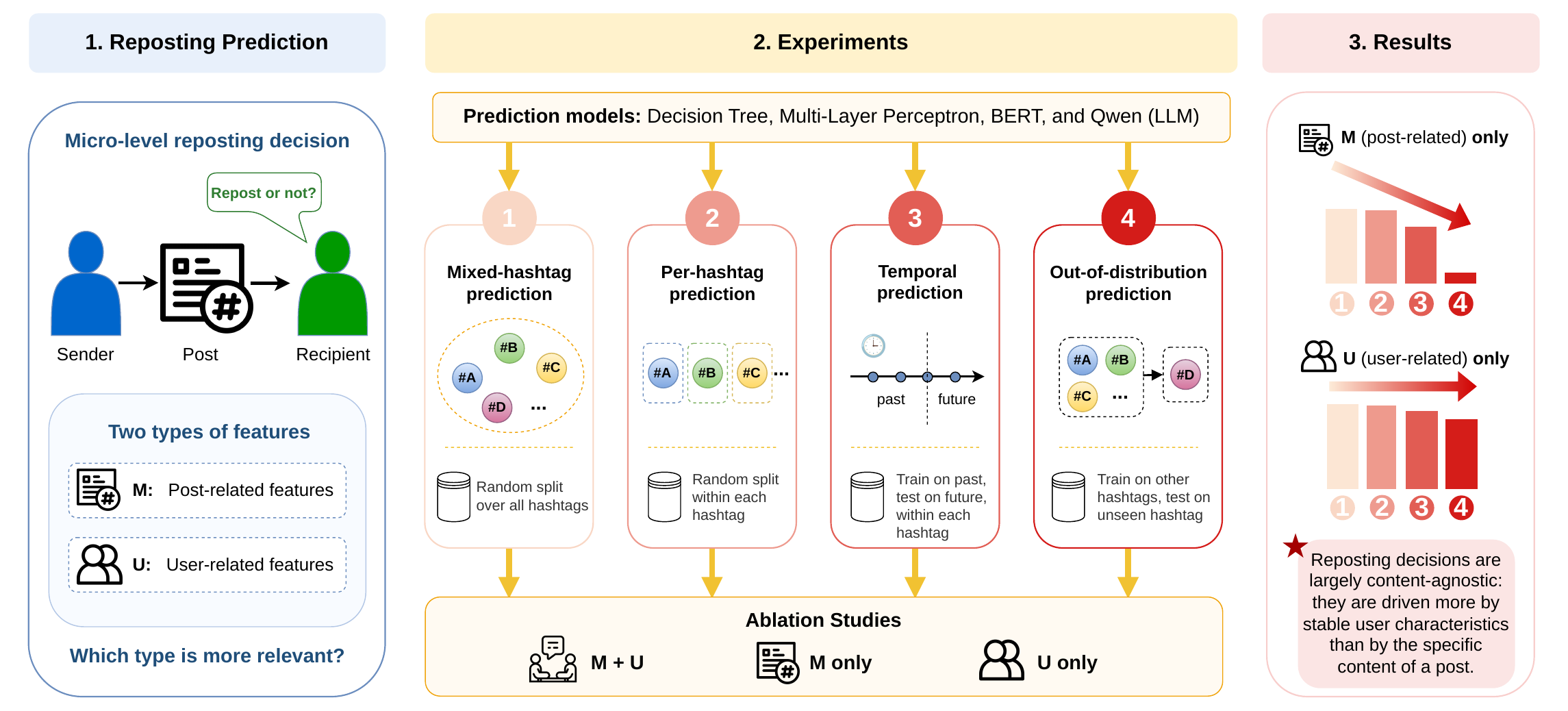}
\caption{Overview of our work. Reposting prediction is studied at the individual level, involving three elements: the Sender, the Post, and the Recipient. A series of experiments are conducted using four different models. For each experiment and model, various settings -- employing different types of features -- are considered for ablation studies.
}
\label{fig:figure_1} 
\end{figure}

\section{Background}

\subsection{Overview of X}

X (formerly Twitter) is a popular online social network, with approximately 600 million monthly active users reported in 2024~\footnote{~\url{https://en.wikipedia.org/wiki/X_(social_network)}}. 
It allows users to publish posts (formerly known as tweets), follow other users to curate their content feed (or `timeline'), and interact with other users through actions, such as `repost' and `like'. 
When a user creates a post, the post is added to the timelines of the user's followers, as well as those who might be interested, as determined by X's recommendation algorithm. 
When a user receives a post, the user can `repost' (formerly known as retweet) the post, `reply' to the post by adding their response, or `quote' the post by adding their comment. 
A user can also `like' a post to signal their endorsement, but this action does not add the post to the timelines of the user's followers.  
The reposting function on X is similar to the `share' function on Facebook. The `share' function on Instagram and the `repost' function on TikTok, however, are different as they forward posts privately to specific users or groups.

\subsection{Related Work}   

There are many related works on various aspects of reposting, but they differ in the prediction target. 
Some works studied posts only, some focused on posts and their senders only, and some considered posts and their recipients only.    
For example,  
\citet{Ma2013OnTwitter, Zheng2022PredictingNetworks} predicted the popularity of a post; 
\citet{li2017deepcas,xu2021casflow} predicted the incremental size of the spreading tree for a given post over time;
\citet{yin2021deep} predicted the time gap between reposts of a given post; 
\citet{wang2022tweet} predicted the probability of a given post 
being reposted (by any user),
and \citet{wang2017topological,islam2018deepdiffuse,cao2021information,yuan2021dyhgcn,yang2021full} predicted the next user who is going to repost a given post based on the sequence of previous reposters of the post and the social relations among these reposters. 

Recent work has increasingly explored the use of large language models (LLMs) for modelling users and their behaviours from textual data.
For example, Retweet-BERT~\cite{jiang2023retweet} uses BERT to extract representations from users' profile descriptions to estimate their political leaning.
Social-LLM~\cite{jiang2025social} extends this work by adding dense layers that incorporate additional user metadata for user-detection tasks.
SoMeR~\cite{guo2025somermultiviewuserrepresentation} learns from users' historical posts for better user representation. 
These studies show that LLMs can be useful for social media user profiling. 
Beyond user modelling, LLMs were also used as user-facing tools for social media content moderation and sharing support. For example, \citet{10.1145/3748699.3749798} use LLMs to identify misinformation and discuss interaction designs that can help users make more informed sharing decisions.

These tasks are closely related to information diffusion, but they do not directly model the sender--post--recipient interaction.
For example, popularity and cascade-size predictions ask whether a post will spread widely; next-reposter prediction asks who may repost next in an observed sequence; post-level reposting prediction asks whether a post will be reposted by anyone; user-profiling studies infer user attributes or interests; content moderation studies focus on platform-level interventions.
In contrast, our goal is to predict whether a specific recipient will repost a specific post from a specific sender and compare which information source supports that event-level prediction.

Prior works on reposting prediction can also be categorised by their input features and the classification algorithm used. 
Specifically, a variety of machine learning (ML) models have been employed, such as Support Vector Machines~\cite{xu2012analyzing,luo2013will}, Decision Trees ~\cite{xu2012analyzing,firdaus2021retweet,zhu2023path,yang2017propagator}, Logistic Regression ~\cite{zhang2013social,quan2018repost}, and factorisation models ~\cite{feng2013retweet,jiang2018retweet,jiang2019retweeting,hoang2016microblogging}. 
Prior studies have reported strong performance from Decision Tree models relative to other ML models in this task~\cite{xu2012analyzing,zhu2023path}. 
ML models usually rely on features extracted from raw data of posts and users. Given the extensive and diverse nature of the features, we elaborate on the taxonomy of these features in a dedicated Section~\ref{sec:features}. 
 
Neural networks have also been used for reposting prediction, including      
the Attention-based Convolutional Neural Network (SUA-ACNN)~\cite{zhang2016retweet}
and the Bidirectional Long Short-Term Memory (Bi-LSTM)~\cite{ma2019hot} models. 
These models learn representations directly from raw data and reduce the need for manual feature extraction.

Reposting prediction has also been studied from the perspective of recommender systems, which predict and rank the likelihood of candidate posts being reposted by a user~\cite{yan2012tweet,feng2013retweet,diaz2014predicting,alawad2016network}. 
These studies used various prediction models, typically considered all three elements, and evaluated their model performance at the ranked-list level.
 
\subsection{Post Data vs User Data}

Prior works on reposting prediction have often relied heavily on post-derived data, including the message text of the post being considered for reposting and the historical posts of users. 
For example, the ML models typically use features extracted from textual post content, while NN models often use the posts data exclusively.

It is known, however, that user-related data, including users' profiles, following lists, past behaviour, interests, and interactions, can have a significant influence on a user's status, actions, and decisions~\cite{jerez2026human,luo2013will,lampos2014predicting,he2021cannot,xu2012analyzing}.
For example, \citet{luo2013will} suggested that a recipient is more likely to repost if they follow the sender, and the two users have shared interests and prior interactions. 
Prior work on recommender systems supports similar observations ~\cite{bobadilla2013recommender,zhang2019deep,wu2022graph}.
Related work on social network exposure has further shown that information exposure is shaped by user network structure and can be algorithmically diversified to mitigate filter bubbles and echo chambers~\cite{matakos2020tell}.

These observations motivate considering social and behavioural context, beyond post content alone, when modelling information diffusion. They also motivate a direct comparison between post-related and user-related features, rather than treating user context as auxiliary metadata.

This comparison is also relevant to language-model-based approaches. Several recent studies ~\cite{jiang2023retweet,jiang2025social,guo2025somermultiviewuserrepresentation} show that LLMs can be useful for social media user profiling.
However, in these studies, language models are mainly applied to textual data, such as profile descriptions, post content, or user historical posts.
Structural user features are typically handled as numerical features or training signals outside the language-model input.
This leaves open the question of whether LLMs can use structured user information and benefit reposting prediction.

\subsection{Out-of-Distribution Evaluation}

Prior works on reposting prediction typically evaluated model performance by randomly splitting the data into training and test sets, so both sets contain data from the same topics~\cite{xu2012analyzing,luo2013will,firdaus2021retweet,zhu2023path,yang2017propagator,zhang2013social,feng2013retweet,jiang2018retweet,jiang2019retweeting,hoang2016microblogging,zhang2016retweet,ma2019hot}. 
This evaluation setup, often referred to as \emph{in-distribution} evaluation~\cite{arjovsky2020out,liu2021towards}, assumes that the training and test data are drawn from the same underlying distribution. 
It is useful for evaluating whether a model can predict reposting behaviour when future cases resemble past cases. 

However, social media platforms continuously produce new topics, and different topics may be associated with different post content and user groups with distinct feature distributions.
As a result, a model trained on previous topics may face a distributional shift when applied to a new topic.
To the best of our knowledge, prior works on reposting prediction have not explicitly evaluated generalisation across different topics under the \emph{out-of-distribution} condition, where a model is trained using data from some topics, characterised by a set of hashtags, and tested using data from other topics. This evaluation assesses a model's capacity to predict reposting behaviour associated with new, previously unseen topics.

\section{Data}

\label{sec:data}
\begin{figure}[!t]
\centering
\includegraphics[width=\textwidth]{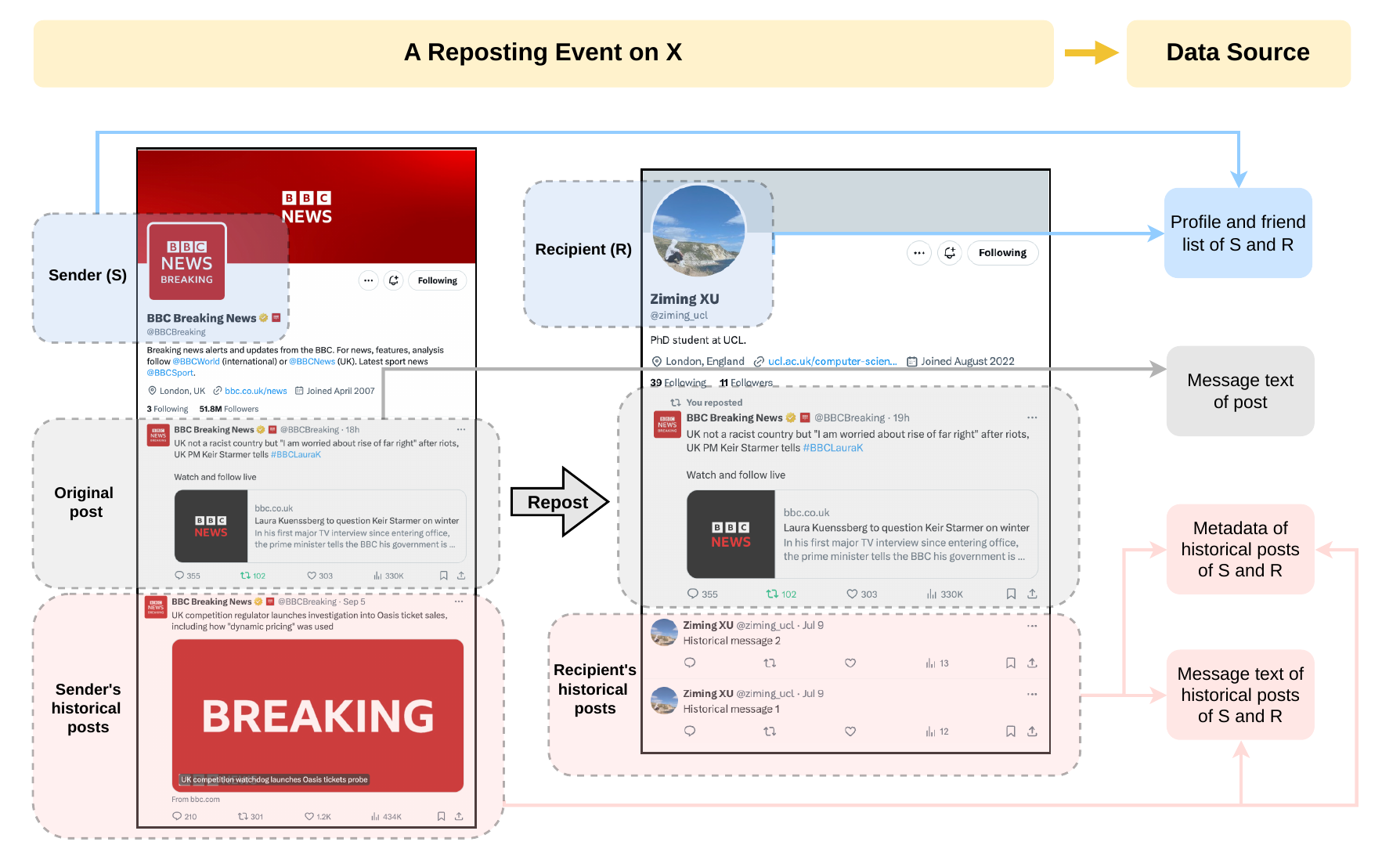}
\caption{Example of a reposting event on X and the corresponding data sources from which different types of features are extracted.
} 
\label{fig:figure_2}
\end{figure}

\begin{table}[!t]
\centering
\small
\setlength{\tabcolsep}{0.8mm}
\renewcommand{\arraystretch}{1.2}
\caption{Types of Features in This Study}
\begin{tabular}{lllll}
\toprule
\textbf{Feature} & {Data} & {Feature} & {Feature}\\
\textbf{Type}\,(\#) &\textbf{Source} & \textbf{Sub-type} (\#) & Example  & Note \\ 
\toprule
\multicolumn{5}{c}{{\textbf{Post-Related} Features}} \\ 
\midrule
\multirow{7}{*}{\shortstack[l]{\textbf{M} (78)\\Message \\ features}} &\multirow{7}{*}{\shortstack[l]{Message text \\ of a post}} & Topic (39) & M-TopicM5 & Likelihood of a post related to topic `family'.\\ 

&& Language (10) & M-Grammar1 & Grammatical correctness of a post. \\ 

&& Readability (11) & M-Readability1& A  measure of difficulty in understanding a post.\\ 

&& Sentiment (5) & M-Sentiment1 & Negative sentiment score of a post.\\ 

&& Emotion (8) & M-Emotion1 & Probability of a post expressing   emotion `anger'.\\ 

&& Hate-Speech (4) & M-Hate1 & A measure of aggressiveness of a post.\\ 

&& Hashtag (1) & M-Hashtag & The presence of a specific hashtag in a post.\\ 

\specialrule{0.08em}{2pt}{2pt}
\multicolumn{5}{c}{{\textbf{User-Related} Features} (of the Sender $U_S$ and the Recipient $U_R$) } \\ 
\midrule
\multirow{4}{*}{\shortstack[l]{\textbf{U-P} (30)\\User Profile \\features}} & \multirow{4}{*}{\shortstack[l]{Profile and \\ friend list \\ of a user } } & Profile of $U_S$ (12) & U-P-S-FollowerNum & The number of followers of the Sender $U_S$.\\ 

&& Profile of $U_R$ (12) & U-P-R-FollowerNum & The number of followers of the Recipient $U_R$.\\ 

& & Network of $U_S$ (3) & U-P-S-FollowR & A binary label indicating $U_S$ follows $U_R$ or not.\\ 

&& Network of $U_R$ (3) & U-P-R-FollowS & A binary label indicating $U_R$ follows $U_S$ or not.\\ 

\midrule
\multirow{5}{*}{\shortstack[l]{\textbf{U-HA} (38) \\ User Historical\\Action features}} &\multirow{5}{*}{\shortstack[l]{Metadata of \\ historical posts \\ of a user}} & Popularity of $U_S$ (4)& U-HA-S-LikedRate & Avg.\,\#\,of ‘Likes' of latest 50 posts of $U_S$. \\ 

& & Popularity of $U_R$ (4) &U-HA-R-LikedRate& Avg.\,\#\,of ‘Likes' of latest 50 posts of $U_R$.\\ 

& & Activity of $U_S$ (7) & U-HA-S-InteractivePer & Percentage of reposts in $U_S$'s latest 50 posts. \\ 

&& Activity of $U_R$(7)  &U-HA-R-InteractivePer& Percentage of reposts in $U_R$'s latest 50 posts.\\ 

&& Interaction (16) & U-HA-RS-Mention & \# of $U_R$'s latest 50 posts that mentioned $U_S$. \\ 

\midrule
\multirow{3}{*}{\shortstack[l]{\textbf{U-HM} (157)\\User Historical \\ Message features}} &\multirow{3}{*}{\shortstack[l]{Message text \\ of historical \\ posts of a user}} & M features of $U_S$ (78) & U-HM-S-Grammar1 & (Each HM feature is averaged over a user's    \\ 

& & M features of $U_R$ (78)  & U-HM-R-Grammar1  & \,\,\,\,\,\,\,\,latest 50 posts.)\\ 

& & Topic Similarity (1) & U-HM-SR-TopicSim  & Topic similarity of latest 50 posts of $U_S$ and $U_R$.\\ 
\bottomrule
\end{tabular}
\label{tab:table_1}
\end{table}

We collected data from X using the former Twitter API under academic access.
We selected the 14 most popular hashtags in the trending list between July 27th and August 14th in 2022, and retrieved all available posts containing these hashtags during the same period around the world. 
We obtained 63,336 posts, including 44,014 reposts (that serve as positive instances).  
These reposts satisfy two conditions: (1)~each repost is created within 24 hours of its original post being posted; (2)~the original post is also present in our data. 
We set a 24-hour window because, according to the study of the half-life of X posts~\cite{pfeffer2023half}, users are unlikely to see a post more than 24 hours after its creation. 
Any reposting after 24 hours is more likely influenced by external factors, such as TV or other social media platforms.
In our X data, reposts account for 78\% of all posts, and replies and quotes account for 2\%.
In this study, we merge replies and quotes into reposts,
because these actions add the post to the timelines of the user's followers and therefore contribute to information diffusion.

These posts (including the reposts) were created by 54,299 unique users, covering all senders and recipients observed in the reposting events. 
For each user, we collected their profile~\footnote{~This includes publicly available account-level metadata, such as follower and followee counts, verification status, account creation time, and profile URL.}, following list, and up to 50 latest historical posts. 
Prior work~\cite{ergu2019predicting,arnoux201725} suggests that 50 historical posts are sufficient to effectively characterise a user.
A total of more than 2.6 million historical posts were collected and later used to construct user-related historical features. All users are anonymised in our data.

A post can contain text and multimedia content, such as pictures, audio, and video. Same as in prior works, in this study, we only consider the message text of a post.
The user data contains three types of user-related data.   
User profile data contains a user's profile and the list of `friends' that the user follows.
User historical action data is the metadata (or public metrics) from a user's latest 50 historical posts. These metadata include: (1)~the type of a post, i.e., original post/repost/quote/reply, (2)~the creation time of a post, (3)~the number of reposts/quotes/replies/likes a post has received, and (4)~any usernames mentioned in a post. These metadata provide useful information about a user's previous behaviours, activity patterns, and interactions with other users.
User historical message text data is the textual content of a user's latest 50 historical posts. These posts characterise a user's historical interests and communication style.

\section{Methodology}

Our methodology has four components.
First, we define the reposting prediction task. Second, we construct post-related and user-related feature groups so that the contribution of each information source can be compared directly.
Third, we build reposting prediction datasets from observed reposting events and non-reposting cases.
Finally, we evaluate the same feature settings across feature-based and language-model-based classifiers.

\subsection{Task Formulation} 

A reposting action involves three elements:
\begin{itemize}
    \item \textbf{The message text of a post ($M$)}. This is the post content being considered for reposting.
    \item \textbf{The sender ($U_S$)}, who created the post. 
    \item \textbf{The recipient ($U_R$)}, i.e.,\,the potential reposter, whose action is to be predicted.
\end{itemize}

In this study, we define reposting prediction as predicting whether a user (recipient) will repost a given post created by another user (sender), considering all three elements involved in a reposting action. 

Specifically, we learn a binary classification model $f$ to predict the label $y$,  
\begin{equation}
\label{eq:problem}
f(M, U_S, U_R) \rightarrow y \in \{0, 1\} \, ,
\end{equation}
where $y=1$ means the recipient will repost the post received from the sender, and $y=0$ otherwise. 

\subsection{Features}
\label{sec:features}

\textit{\textbf{Message Features (M).}} 
From the message text of a post, we extract 78 post-related message features, denoted as M, which are used in prior works for reposting prediction.
These features characterise the post being considered for reposting from different perspectives.
As shown in Table\,\ref{tab:table_1}, we divide them into sub-types, including topic
~\cite{antypas2022twitter,hong2011predicting,martin2016exploring,zhu2023path,firdaus2021retweet,hoang2016microblogging}, language style~\cite{barbieri2020tweeteval,feng2013retweet,khabiri2009analyzing,chen2020event}, readability~\cite{khabiri2009analyzing}, sentiment~\cite{hutto2014vader,robertson2023negativity,zhao2018comparative,firdaus2021retweet,upadhyaya2022spotting}, emotion~\cite{paletz2023emotional,robertson2023negativity,wang2021effect,firdaus2021retweet,upadhyaya2022spotting}, hate speech
~\cite{mousavi2022effective}, and hashtag. 
These features are calculated using publicly available language models and tools (see Appendix~\ref{app:features}).
In addition, the hashtag feature is recorded as a categorical label~\cite{feng2013retweet,firdaus2021retweet}.

\textit{\textbf{User Profile Features (U-P).}} 
These features are extracted from user profile data. 
They include the number of followers/followees of a user~\cite{paletz2023emotional,martin2016exploring,zhou2020realistic,chen2020event,upadhyaya2022spotting}, the number of lists that a user is included in~\cite{luo2013will,xu2012analyzing}, 
the number of posts a user has created since its registration~\cite{luo2013will,huberman2008social,martin2016exploring,upadhyaya2022spotting},
whether a user is verified~\cite{feng2013retweet,xu2012analyzing}, whether a URL is provided in a user's profile~\cite{bakshy2011everyone}, 
a user's indegree~\cite{zhu2023path} and LeaderRank score~\cite{zhu2023path}, and whether a user is following another user~\cite{xu2012analyzing,upadhyaya2022spotting}. 
Together, these features describe properties of the sender and recipient, as well as following relations between them, which may provide useful predictive signals for reposting prediction.

\textit{\textbf{User Historical Action Features (U-HA).}}
These features are extracted from user historical action data. 
They include, for example, the percentage of a user's historical posts that are reposts~\cite{feng2013retweet,zhu2023path}, the average time interval between a user's posts~\cite{feng2013retweet,upadhyaya2022spotting}, and the ratio of a sender's historical posts that have mentioned a recipient~\cite{luo2013will,xu2012analyzing}.  
These features are used to represent historical activity patterns and interaction signals.

\textit{\textbf{User Historical Message Features (U-HM).}} 
These features are extracted from user historical message text data.
They include the aggregated Message features (M) of a sender and those of a recipient, as well as an additional feature measuring topic similarity between the two users~\cite{zhu2023path,quan2018repost}.
These features describe historical interests and communication patterns, such as topic interests and language style, inferred from users' historical posts.

More details of the features are provided in Appendix ~\ref{app:features}.

\subsection{Dataset Construction}
\label{3Datasets}

In this study, we define a positive instance as a reposting event $(M_h, U_{S,t_s}, U_{R,t_r})$, 
where the (original) post~$M_h$ with a hashtag $h$ was created by a sender~$U_S$ at $t_s$, and was reposted by a recipient~$U_R$ at time $t_r$ (within 24 hours since $t_s$).   
Correspondingly, a negative instance is $(M^*_h, U^*_{S,t_s}, U_{R,t_r})$, where the same recipient $U_R$ did not repost another post $M^*_h$ with the same hashtag $h$ created by sender $U^*_S$ (where it is possible $U^*_S = U_S$). 
To make the negative instance more comparable with the corresponding positive instance, we require the negative post to contain the same hashtag as the positive post and to be created in the same 24-hour window.
This design makes the negative post a same-topic, contemporaneous alternative rather than an arbitrary non-reposted post.

Our main reposting-prediction dataset uses a 1:5 ratio of positive to negative instances.
For each positive instance, the five most similar negative instances are selected from our X data.
Specifically, the post and the sender in each negative instance are selected to be most similar to those in the corresponding positive instance, i.e., the cosine distance between the feature vectors (containing all features of the post and the sender) of the positive instance and the chosen negative instance is among the five smallest across candidate negative instances.
This 1:5 ratio is typically used in reposting-prediction experiments, such as in \citet{zhang2016retweet} and \citet{ma2019hot}.

We also construct two supplementary datasets with 1:1 and 1:10 positive-to-negative ratios for robustness checks.
For the 1:1 dataset, one most similar negative instance is selected for each positive instance.
For the 1:10 dataset, in addition to the five most similar negative instances, five more negative instances are randomly chosen from our X data.
Results on these supplementary datasets are reported in Appendix~\ref{app:results_1_1_1_10}.

Note that in all experiments in this study, we remove any instance from the training set if its sender--recipient pair also appears in the test set. 
Thus, training and test instances always have different sender--recipient pairs.  
This is to prevent information leakage and ensure a fair and rigorous evaluation.

\subsection{Models}

We use four models to test whether the relative value of post-related and user-related information is model-specific or stable across different classifiers. 
The models include a decision tree (DT), a multi-layer perceptron (MLP), BERT, and Qwen.
The purpose of this comparison is not to claim that one model family is inherently superior, but to examine whether the feature-level pattern is consistent across feature-based and language-model-based classifiers.

We follow standard practices for each model while keeping the settings of three information sources conceptually aligned.
For DT and MLP, the inputs are engineered feature vectors.
For BERT and Qwen, the inputs are textual representations: message text of the post for M, serialised user-related features for U, and their concatenation for ALL.
 
\textit{\textbf{Decision Tree (DT)}.}
The DT model remains a competitive choice for reposting prediction~\cite{zhu2023path, xu2012analyzing}. 
In this study, we use a DT model to examine how prediction performance changes when different engineered feature groups are provided as input. 
Specifically, DT-M uses post-related features M, DT-U uses user-related features U, including U-P, U-HA, and U-HM, and DT-ALL uses the concatenation of M and U.

We implement the DT model with XGBoost and tune hyperparameters for each data split of each dataset. The most frequently used hyperparameters include the maximum depth of 8, the learning rate of 0.3, the estimators of 100, the minimum child weight of 1, the subsample ratio of 1.0, and the positive class weight of 1.0. For more details, please refer to Appendix~\ref{app:DT_parameters}.

\textit{\textbf{Multi-Layer Perceptron (MLP).}}
Neural Networks
have become popular choices for reposting prediction.
We use a feature-based multi-layer perceptron (MLP), which is directly comparable with the DT model because both models receive the same engineered feature vectors under the M, U, and ALL settings.

The MLP has two hidden layers with 256 and 128 units, respectively, and uses a dropout rate of $0.1$. It is trained with the \textit{Adam} optimiser, the cross-entropy loss, the learning rate of $0.001$, the weight decay of $1e-5$, the batch size of $512$, the maximum number of epochs of $50$, and early stopping on validation loss with a patience of $10$.

\textit{\textbf{BERT.}}
We also study BERT~\cite{devlin2019bert}, which is a widely studied large language model (LLM). 
Unlike DT and MLP, which use engineered feature vectors, BERT receives textual input.
Specifically, BERT-M takes the post text as input. 
BERT-U takes serialised user-related features as input. The user-related features are converted into text using the serialisation strategy in \citet{hegselmann2023tabllm}, so that BERT can process structured user information as text. An example input is provided in Appendix~\ref{app:example_serialisations}.
BERT-ALL takes the concatenation of the post content and the serialised user-related features as input.
The model generates embeddings for the input text and uses a classification layer to predict whether the recipient will repost the post from the sender.

We used the recommended fine-tuning settings from \citet{devlin2019bert}. We configure BERT-base with the \textit{AdamW} optimiser, the learning rate of $2e-5$, the batch size of $32$, the dropout rate of $0.1$, the maximum number of epochs of $5$, and early stopping on validation loss with patience of $2$. Cross-entropy loss is used as the objective function.

\textit{\textbf{Qwen}.}
We further include Qwen~\cite{yang2025qwen3technicalreport} as a recent advanced open-weight large language model (LLM).
As with BERT, the input depends on the setting:
Qwen-M uses the post content.
Qwen-U uses serialised user-related features.
Qwen-ALL uses their concatenation.

In our implementation, we use \texttt{Qwen3-0.6B} as a lightweight representative of the Qwen3 model family.
In preliminary experiments, \texttt{Qwen3-4B} and \texttt{Qwen3-1.7B} produced similar performance (see Appendix~\ref{app:qwen_varients}), so the 0.6B model is sufficient for our comparison.
Using the smaller model also facilitates repeated fine-tuning under multiple feature settings and evaluation splits, and improves the reproducibility of our experiments.

We perform supervised fine-tuning (SFT) using LoRA~\cite{hu2022lora} with rank $8$, alpha $16$, and target modules set to all LoRA-supported modules.
We use the recommended fine-tuning settings from \citet{zheng-etal-2024-llamafactory} as closely as practical, except that the batch size is adjusted for computational efficiency.
We use the learning rate of $1e-4$, the batch size of $4$, gradient accumulation steps of $8$, the maximum number of epochs of $3$, the warm-up ratio of $0.1$, and a cosine learning-rate scheduler. The best checkpoint is selected according to validation loss.
We use the probability assigned to token \texttt{1} as the predicted reposting probability. The system prompt is provided in Appendix~\ref{app:qwen_prompt}.

\section{Experiments and Results}

We evaluate reposting prediction with F1 score through four experiments arranged from in-distribution evaluation to out-of-distribution generalisation.
Across all experiments, we compare three feature settings: M (post-related message features), U (User-related features), and ALL (their combination).
This design tests whether reposting prediction is better supported by what is posted, by information about the sender and recipient, or by both sources together.

\subsection{In-Distribution Mixed-Hashtag Prediction}

\textit{\textbf{Experimental Setting}.}
In this experiment, we pool data from all hashtags and perform Monte Carlo cross-validation 10 times, each time randomly splitting the data 63/7/30 into training, validation, and test sets. 
We report the mean and standard deviation of the F1 scores across the 10 Monte Carlo cross-validation runs.
Due to the substantially higher computational cost of fine-tuning Qwen, each Qwen-based model is evaluated over 5 runs rather than 10 in this mixed-hashtag prediction, as well as in the per-hashtag and out-of-distribution predictions conducted separately for each hashtag. This provides multiple independent runs for estimating the stability of the results while keeping the experiments computationally feasible.

\textit{\textbf{Results}.}
As shown in Table \ref{tab:table_2}, models using ALL features, i.e., DT-ALL, MLP-ALL, BERT-ALL, and Qwen-ALL, outperform their corresponding U-only and M-only variants.
Among models with reduced feature sets, U consistently outperforms M across the four models.
Both paired t-tests and Wilcoxon signed-rank tests confirm that the above performance rankings are statistically significant (with all values of \textit{p} $<$ 0.05).
These results show that post-related features support reposting prediction under in-distribution evaluation, while user-related features provide additional predictive signal across model classes.

\begin{table}[!t]
\centering
\renewcommand{\arraystretch}{1.2}
\caption{\textbf{In-Distribution} Prediction Results (F1 $\pm$ Std.) using Decision Tree (DT), Multi-Layer Perceptron (MLP), BERT, and Qwen
}
\begin{tabular}{cc|ccc}
\hline
&  & \multicolumn{3}{c}{Feature Setting} \\ 
Experiment & Model & \cellcolor[HTML]{FFF8DC} \makebox[2cm][c]{\textbf{ALL} (U + M)} & \cellcolor[HTML]{FFF8DC} \makebox[2cm][c]{\textbf{U} (user-related)} & \cellcolor[HTML]{FFF8DC} \makebox[2cm][c]{\textbf{M} (post-related)} \\ 
\hline
& DT 
& \cellcolor[HTML]{EAF3FD}\textbf{0.884}\scriptsize{$\pm$0.002} 
& 0.852\scriptsize{$\pm$0.005} 
& 0.758\scriptsize{$\pm$0.002} \\ 

{\bf Mixed-Hashtag} & MLP
& \cellcolor[HTML]{EAF3FD}\textbf{0.847}\scriptsize{$\pm$0.004} 
& 0.822\scriptsize{$\pm$0.003} 
& 0.715\scriptsize{$\pm$0.003} \\ 

Prediction & BERT
& \cellcolor[HTML]{EAF3FD}\textbf{0.853}\scriptsize{$\pm$0.017}
& 0.824\scriptsize{$\pm$0.013}
& 0.740\scriptsize{$\pm$0.010} \\

& Qwen
& \cellcolor[HTML]{EAF3FD}\textbf{0.871}\scriptsize{$\pm$0.002} & 0.848\scriptsize{$\pm$0.003} & 0.746\scriptsize{$\pm$0.004} \\

\hline
& DT 
& \cellcolor[HTML]{EAF3FD}\textbf{0.868}\scriptsize{$\pm$0.040}
& 0.841\scriptsize{$\pm$0.069} 
& 0.742\scriptsize{$\pm$0.073} \\ 

{\bf Per-Hashtag} & MLP
& \cellcolor[HTML]{EAF3FD}\textbf{0.829}\scriptsize{$\pm$0.049}
& 0.805\scriptsize{$\pm$0.067} 
& 0.728\scriptsize{$\pm$0.080} \\ 
 
Prediction & BERT
& \cellcolor[HTML]{EAF3FD}\textbf{0.786}\scriptsize{$\pm$0.060}
& 0.776\scriptsize{$\pm$0.090}
& 0.725\scriptsize{$\pm$0.078} \\

& Qwen
& \cellcolor[HTML]{EAF3FD} \textbf{0.826}\scriptsize{$\pm$0.055}
& 0.795\scriptsize{$\pm$0.078}
& 0.716\scriptsize{$\pm$0.085} \\ 

\hline
& DT 
& \cellcolor[HTML]{EAF3FD}\textbf{0.815}\scriptsize{$\pm$0.091}
& 0.788\scriptsize{$\pm$0.107}
& 0.579\scriptsize{$\pm$0.212} \\ 

{\bf Temporal} & MLP
& \cellcolor[HTML]{EAF3FD}\textbf{0.760}\scriptsize{$\pm$0.101}
& 0.753\scriptsize{$\pm$0.105}
& 0.554\scriptsize{$\pm$0.210} \\ 

Prediction & BERT 
& 0.592\scriptsize{$\pm$0.202}
& \cellcolor[HTML]{EAF3FD}\textbf{0.676}\scriptsize{$\pm$0.160}
& 0.530\scriptsize{$\pm$0.248} \\ 

& Qwen
& 0.733\scriptsize{$\pm$0.130} 
& \cellcolor[HTML]{EAF3FD}\textbf{0.740}\scriptsize{$\pm$0.114}
& 0.519\scriptsize{$\pm$0.244}  \\ 

\hline
\multicolumn{5}{p{0.7\linewidth}}{\footnotesize Note: We {\setlength{\fboxsep}{1pt}\colorbox[HTML]{EAF3FD}{\textbf{highlight}}} the best performance of the three feature settings for each experiment and model.}
\end{tabular}
\label{tab:table_2}
\end{table}

\begin{table}[!t]
\centering
\renewcommand{\arraystretch}{1.2}
\caption{
\textbf{Top 20 Important Features} for the In-Distribution Mixed-Hashtag Prediction using the Decision Tree (DT) model
}
\begin{tabular}{lcllc}
\toprule
\textbf{Feature name} & \textbf{Rank}  & \textbf{Feature type}  & \textbf{Subtype} & \textbf{Importance Value} \\
\midrule
U-P-R-FollowS& 1 & U-P & network& 0.229\\ 
U-HA-S-RetweetedRate& 2 & U-HA& popularity& 0.047\\ 
U-HA-S-LikedRate& 3 & U-HA& popularity& 0.045\\ 
U-HA-S-RepliedRate& 4 & U-HA& popularity& 0.042 \\ 
U-HA-RS-MentionPer& 5 & U-HA& interaction & 0.015 \\ 
M-TopicLDA7 $^{1}$ & 6 & M& topic    & 0.013 \\ 
U-HA-S-InteractivePer& 7 & U-HA& activity& 0.013 \\ 
U-HA-S-TweetNum& 8 & U-HA& activity& 0.011 \\ 
U-HA-S-RetweetPercent& 9 & U-HA& activity& 0.009 \\ 
U-P-S-FollowR& 10 & U-P& network& 0.008 \\ 
U-HA-RS-RepostLatency& 11 & U-HA& interaction & 0.008 \\ 
U-HM-S-Readability7& 12 & U-HM & readability& 0.008 \\ 
M-TopicLDA4& 13 & M& topic    & 0.008 \\ 
M-TopicLDA3& 14 & M& topic    & 0.008 \\ 
U-HM-S-TopicM7& 15 & U-HM& topic    & 0.008 \\ 
U-HM-S-TopicM11& 16 & U-HM& topic    & 0.007 \\ 
U-HM-S-TopicM19& 17 & U-HM & topic    & 0.007 \\ 
U-HA-RS-Mention& 18 & U-HA& interaction & 0.007 \\ 
U-HA-S-TweetPercent& 19 & U-HA& activity& 0.007 \\ 
U-P-S-Indegree& 20 & U-P& network& 0.007 \\ 
\midrule
\multicolumn{5}{l}{\# of features in each type: U-HA(10), U-HM(4), U-P(3), M(3).} \\ 
\bottomrule
\multicolumn{5}{p{0.7\linewidth}}{\footnotesize $^{1}$M-TopicLDA7 represents topics related to energy crisis, Brexit, and health issues. See full details of features in the Appendix~\ref{app:features}. }
\end{tabular}
\label{tab:table_3}
\end{table}

\textit{\textbf{Feature Importance}.}
The feature importance value represents the relative contribution of a feature to the decision tree model's predictive performance. Specifically, it is computed as the reduction in the loss function contributed by each feature when splitting nodes, with larger reductions indicating higher importance.
We obtained the feature importance values for all 303 features by training the DT model under the ALL setting using all (instead of 70\%) data in the 1:5 dataset. The feature importance values are normalised such that the sum for all features equals one.
Fewer than 1\% of feature pairs have a Pearson correlation coefficient higher than 0.7, suggesting limited high pairwise collinearity among the 303 features and most features contribute to the prediction~\cite{dormann2013collinearity}. 

Table \ref{tab:table_3} shows the top 20 most important features.
It is notable that the majority of the top 20 features are user-related features, including 10 user historical action (U-HA) features, 4 user historical message (U-HM) features, and 3 user profile (U-P) features. 
The most important message (M) feature is merely ranked 6th.   
This feature importance analysis is consistent with and highlights the stronger predictive performance of user-related features.

\subsection{In-Distribution Per-Hashtag Prediction}

\textit{\textbf{Experimental Setting}.}
In this experiment, we conduct predictions separately for each of the 14 hashtags. 
For each hashtag, we perform the same Monte Carlo cross-validation as in the mixed-hashtag prediction, but only using data from that individual hashtag. 
This tests whether the relative contribution of M, U, and ALL features remains stable across individual topics.
To summarise performance across 14 hashtag-specific experiments, we first compute the mean $\mu_i$ and standard deviation $\sigma_i$ of the F1 scores across the 10 cross-validation folds for each hashtag $i$.
We then compute the overall mean across the 14 hashtags 
and the corresponding overall standard deviation,
by treating the results from the 14 hashtags as equally weighted components of a mixture distribution.

\textit{\textbf{Results}.}
Table \ref{tab:table_2} presents the average prediction results across individual hashtags.
The detailed results for each hashtag are provided in Appendix~\ref{app:per_hashtag_results}.
The results are broadly consistent with those observed from the mixed-hashtag prediction, with minor variations across individual hashtags.
Specifically, across model families, models using ALL or U outperform models using M alone.
Both paired t-tests and Wilcoxon signed-rank tests confirm that the above performance rankings are statistically significant (with all values of \textit{p} $<$ 0.05).
These findings indicate that the relative effectiveness of user-related and post-related features is stable across different topics.

The difference between mixed-hashtag prediction and per-hashtag prediction also shows that the scope of the training data can affect performance.
The mixed hashtag scope, which leverages data across all 14 hashtags, generally produces better performance than the per-hashtag scope. This suggests benefits from increased diversity and volume of the training data. 

\subsection{In-Distribution Temporal Prediction}

\textit{\textbf{Experimental Setting}.}
In this experiment, we evaluate whether models trained on earlier reposting events can predict later reposting events within the same hashtag.
For each hashtag, we divide the dataset into 13 time windows with equal numbers of positive instances (and their corresponding negative instances), ordered by the timestamp of reposting events. 
The first three windows are used for training, and the next window is used for testing. 
We then roll the split forward by one window at each step, producing 10 sequential predictions for each hashtag.
For each sequential prediction, the latest 10\% of the training data is used for validation.
The mean and standard deviation of F1 scores are computed across the 10 sequential predictions for each hashtag, and then aggregated across the 14 hashtags using the same procedure as in per-hashtag prediction.
This temporal splitting complements random splitting as it respects temporal order and tests prediction on future instances rather than randomly held-out instances.

\begin{figure}[!t]
\centering
\includegraphics[width=\textwidth]{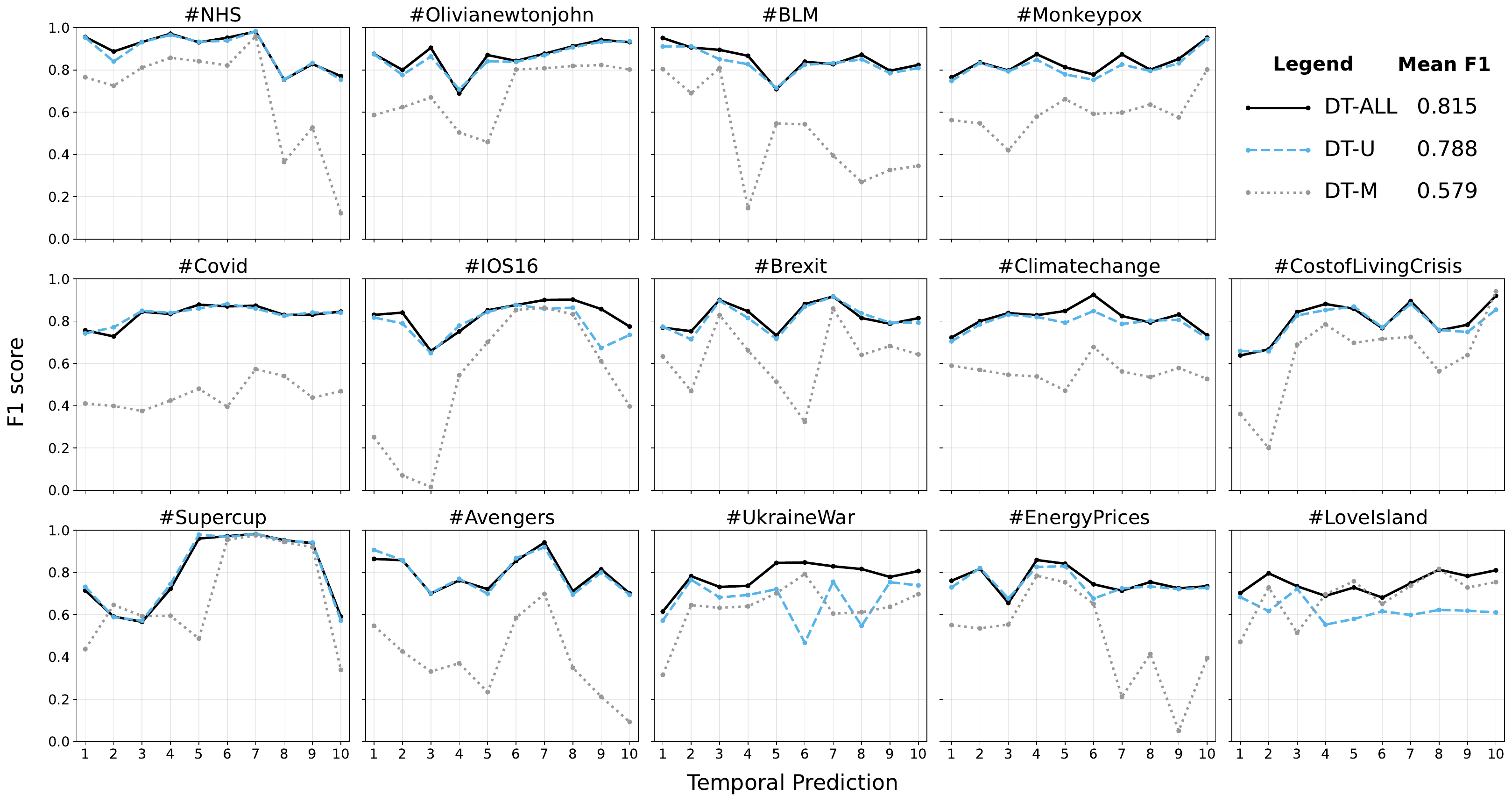}
\caption{In-Distribution \textbf{Temporal Prediction} for 14 individual hashtags using the Decision Tree model with three settings, i.e., using All features (DT-ALL), using User-related features only (DT-U), and using post-related Message features only (DT-M). 
For each hashtag, data are divided into 13 temporal windows. 
Starting from window 4, a model is trained on the previous three windows and tested on the current window. A sequence of 10 steps of temporal predictions are produced for each setting.}
\label{fig:figure_3} 
\end{figure}

\textit{\textbf{Results}.}
The average temporal results are shown in Table \ref{tab:table_2}, and the detailed results for each hashtag are provided in Appendix~\ref{app:temporal_results}.
The results show a clear performance drop compared with the random-split experiments, which is expected because temporal splitting evaluates prediction under temporal shift, which is more realistic and challenging.
Nevertheless, the main pattern remains consistent: models using ALL or U features generally outperform models using M features alone. 
Both paired t-tests and Wilcoxon signed-rank tests confirm that the above performance rankings are statistically significant (with all values of \textit{p} $<$ 0.05).
This suggests that user-related features provide useful predictive signals not only under random splits, but also when predicting later reposting behaviour within the same topic.

We also provide a visualisation of F1 scores across sequential predictions for each hashtag in Figure \ref{fig:figure_3}.
The figure compares DT-ALL, DT-U, and DT-M, allowing us to examine whether the relative performance of post-related and user-related features remains stable over time within each hashtag.
Overall, the temporal patterns are broadly consistent with the aggregate results: DT-ALL and DT-U tend to outperform DT-M across most hashtags and sliding steps, although the magnitude of the gap varies.

\textit{\textbf{Sliding Windows vs. Accumulated Windows}.}
We use sliding windows as the main temporal split strategy because this keeps the size of the training data comparable across sequential predictions.
We also evaluate an accumulated-window split, where all previous windows are used for training. For example, when predicting time window 6, the accumulated-window split uses windows 1--5 for training, whereas the sliding-window split uses windows 3--5.
The two split strategies produce very similar results, so the main temporal findings are not sensitive to this choice of training time windows.
The detailed comparison between sliding and accumulated windows is in Appendix~\ref{app:accum_time_windows}.

\subsection{Out-of-Distribution Prediction}

\begin{table}[!t]
\centering
\renewcommand{\arraystretch}{1.2}
\caption{
\textbf{Out-of-Distribution} Prediction Results }
\begin{tabular}{r|ccc|ccc}
\toprule
\textbf{Model:} & \multicolumn{3}{c|}{{\textbf{DT}}} & \multicolumn{3}{c}{{\textbf{MLP}}} \\ 
\cellcolor[HTML]{FFF8DC}\textbf{Feature Setting:} & \cellcolor[HTML]{FFF8DC}\textbf{ALL} & \cellcolor[HTML]{FFF8DC}\textbf{U-only} & \cellcolor[HTML]{FFF8DC}\textbf{M-only} & \cellcolor[HTML]{FFF8DC}\textbf{ALL} & \cellcolor[HTML]{FFF8DC}\textbf{U-only} & \cellcolor[HTML]{FFF8DC}\textbf{M-only} \\  
\midrule
Hashtag & \multicolumn{6}{c}{{F1 Score ($\mu$ $\pm$ $\sigma$)}} \\ 
\hline
\texttt{\#BLM} 
& 0.812\scriptsize{$\pm$0.015} & \cellcolor[HTML]{EAF3FD}\textbf{0.832}\scriptsize{$\pm$0.007}& 0.347\scriptsize{$\pm$0.232}
& 0.741\scriptsize{$\pm$0.033}& \cellcolor[HTML]{EAF3FD}\textbf{0.773}\scriptsize{$\pm$0.021}& 0.072\scriptsize{$\pm$0.053}\\  
\texttt{\#Covid} 
& \cellcolor[HTML]{EAF3FD}\textbf{0.809}\scriptsize{$\pm$0.010} & 0.788\scriptsize{$\pm$0.011}& 0.095\scriptsize{$\pm$0.045}
& \cellcolor[HTML]{EAF3FD}\textbf{0.743}\scriptsize{$\pm$0.008}& 0.732\scriptsize{$\pm$0.012}& 0.090\scriptsize{$\pm$0.040}\\  
\texttt{\#NHS} 
& 0.746\scriptsize{$\pm$0.065} & \cellcolor[HTML]{EAF3FD}\textbf{0.769}\scriptsize{$\pm$0.061}& 0.143\scriptsize{$\pm$0.141}
& 0.619\scriptsize{$\pm$0.064}& \cellcolor[HTML]{EAF3FD}\textbf{0.720}\scriptsize{$\pm$0.055}& 0.100\scriptsize{$\pm$0.020}\\  
\texttt{\#Monkeypox} 
& \cellcolor[HTML]{EAF3FD}\textbf{0.735}\scriptsize{$\pm$0.010} & 0.692\scriptsize{$\pm$0.048}& 0.133\scriptsize{$\pm$0.026}& 0.694\scriptsize{$\pm$0.006}& \cellcolor[HTML]{EAF3FD}\textbf{0.698}\scriptsize{$\pm$0.009}& 0.037\scriptsize{$\pm$0.017}\\  
\texttt{\#Climatechange} 
& 0.733\scriptsize{$\pm$0.057} & \cellcolor[HTML]{EAF3FD}\textbf{0.736}\scriptsize{$\pm$0.022}& 0.127\scriptsize{$\pm$0.112}
& 0.735\scriptsize{$\pm$0.013}& \cellcolor[HTML]{EAF3FD}\textbf{0.748}\scriptsize{$\pm$0.009}& 0.085\scriptsize{$\pm$0.013}\\  
\texttt{\#Brexit} 
& 0.729\scriptsize{$\pm$0.014} & \cellcolor[HTML]{EAF3FD}\textbf{0.745}\scriptsize{$\pm$0.014}& 0.054\scriptsize{$\pm$0.033}
& 0.674\scriptsize{$\pm$0.032}& \cellcolor[HTML]{EAF3FD}\textbf{0.691}\scriptsize{$\pm$0.016}& 0.102\scriptsize{$\pm$0.051}\\  
\texttt{\#IOS16} 
& 0.723\scriptsize{$\pm$0.025} & \cellcolor[HTML]{EAF3FD}\textbf{0.749}\scriptsize{$\pm$0.020}& 0.085\scriptsize{$\pm$0.124}
& 0.657\scriptsize{$\pm$0.032}& \cellcolor[HTML]{EAF3FD}\textbf{0.722}\scriptsize{$\pm$0.014}& 0.053\scriptsize{$\pm$0.044}\\  
\texttt{\#Supercup} 
& 0.708\scriptsize{$\pm$0.019} & \cellcolor[HTML]{EAF3FD}\textbf{0.741}\scriptsize{$\pm$0.026}& 0.201\scriptsize{$\pm$0.177}
& \cellcolor[HTML]{EAF3FD}\textbf{0.685}\scriptsize{$\pm$0.013}& 0.671\scriptsize{$\pm$0.023}& 0.006\scriptsize{$\pm$0.006}\\  
\texttt{\#Olivianewtonjohn} 
& 0.668\scriptsize{$\pm$0.047} & \cellcolor[HTML]{EAF3FD}\textbf{0.748}\scriptsize{$\pm$0.033}& 0.084\scriptsize{$\pm$0.082}
& 0.463\scriptsize{$\pm$0.053} & \cellcolor[HTML]{EAF3FD}\textbf{0.699}\scriptsize{$\pm$0.029}& 0.169\scriptsize{$\pm$0.073}\\  
\texttt{\#CostofLivingCrisis} 
& 0.652\scriptsize{$\pm$0.027} & \cellcolor[HTML]{EAF3FD}\textbf{0.667}\scriptsize{$\pm$0.025}& 0.036\scriptsize{$\pm$0.022}
& 0.530\scriptsize{$\pm$0.048} & \cellcolor[HTML]{EAF3FD}\textbf{0.641}\scriptsize{$\pm$0.007}& 0.052\scriptsize{$\pm$0.027}\\  
\texttt{\#Avengers} 
& 0.628\scriptsize{$\pm$0.023} & \cellcolor[HTML]{EAF3FD}\textbf{0.665}\scriptsize{$\pm$0.026} & 0.063\scriptsize{$\pm$0.065}
& 0.535\scriptsize{$\pm$0.028}& \cellcolor[HTML]{EAF3FD}\textbf{0.595}\scriptsize{$\pm$0.033}& 0.099\scriptsize{$\pm$0.119}\\  
\texttt{\#UkraineWar} 
& \cellcolor[HTML]{EAF3FD}\textbf{0.618}\scriptsize{$\pm$0.025} & 0.562\scriptsize{$\pm$0.036}& 0.066\scriptsize{$\pm$0.039}
& \cellcolor[HTML]{EAF3FD}\textbf{0.636}\scriptsize{$\pm$0.018}& 0.630\scriptsize{$\pm$0.016}& 0.173\scriptsize{$\pm$0.066}\\  
\texttt{\#EnergyPrices} 
& \cellcolor[HTML]{EAF3FD}\textbf{0.599}\scriptsize{$\pm$0.032} & 0.598\scriptsize{$\pm$0.029}& 0.164\scriptsize{$\pm$0.033}
& 0.531\scriptsize{$\pm$0.013}& \cellcolor[HTML]{EAF3FD}\textbf{0.543}\scriptsize{$\pm$0.011}& 0.085\scriptsize{$\pm$0.036}\\  
\texttt{\#LoveIsland} 
& \cellcolor[HTML]{EAF3FD}\textbf{0.594}\scriptsize{$\pm$0.022} & 0.574\scriptsize{$\pm$0.023}& 0.045\scriptsize{$\pm$0.048}
& 0.525\scriptsize{$\pm$0.015} & \cellcolor[HTML]{EAF3FD}\textbf{0.535}\scriptsize{$\pm$0.013}& 0.044\scriptsize{$\pm$0.027}\\  
\hline
\multirow{2}{*}{ } & \multicolumn{6}{c}{\textbf{Overall} F1 Score ($\bar\mu \pm \bar\sigma$)} \\
& 0.697\scriptsize{$\pm$0.076} & \cellcolor[HTML]{EAF3FD}\textbf{0.705}\scriptsize{$\pm$0.084}& 0.117\scriptsize{$\pm$0.131}
& 0.626\scriptsize{$\pm$0.095}& \cellcolor[HTML]{EAF3FD}\textbf{0.671}\scriptsize{$\pm$0.074}& 0.083\scriptsize{$\pm$0.068}\\  
\hline 
\hline
\textbf{Model:} & \multicolumn{3}{c|}{\textbf{BERT}} & \multicolumn{3}{c}{\textbf{Qwen}}\\ 
\cellcolor[HTML]{FFF8DC}\textbf{Feature Setting:} & \cellcolor[HTML]{FFF8DC}\textbf{ALL} & \cellcolor[HTML]{FFF8DC}\textbf{U-only} & \cellcolor[HTML]{FFF8DC}\textbf{M-only} & \cellcolor[HTML]{FFF8DC}\textbf{ALL} & \cellcolor[HTML]{FFF8DC}\textbf{U-only} & \cellcolor[HTML]{FFF8DC}\textbf{M-only} \\    
\toprule
Hashtag & \multicolumn{6}{c}{{F1 Score ($\mu$ $\pm$ $\sigma$)}} \\ 
\hline
\texttt{\#BLM} 
& 0.765\scriptsize{$\pm$0.035}& \cellcolor[HTML]{EAF3FD}\textbf{0.787}\scriptsize{$\pm$0.049}& 0.123\scriptsize{$\pm$0.142} & 0.804\scriptsize{$\pm$0.033}& \cellcolor[HTML]{EAF3FD}\textbf{0.808}\scriptsize{$\pm$0.015}& 0.277\scriptsize{$\pm$0.235}\\  
\texttt{\#Covid} 
& \cellcolor[HTML]{EAF3FD}\textbf{0.746}\scriptsize{$\pm$0.018}& 0.742\scriptsize{$\pm$0.035}& 0.063\scriptsize{$\pm$0.051} & \cellcolor[HTML]{EAF3FD}\textbf{0.769}\scriptsize{$\pm$0.015}& 0.761\scriptsize{$\pm$0.014}& 0.068\scriptsize{$\pm$0.032}\\  
\texttt{\#NHS} 
& \cellcolor[HTML]{EAF3FD}\textbf{0.617}\scriptsize{$\pm$0.108}& 0.581\scriptsize{$\pm$0.079}& 0.190\scriptsize{$\pm$0.212} & 0.646\scriptsize{$\pm$0.130}& \cellcolor[HTML]{EAF3FD}\textbf{0.791}\scriptsize{$\pm$0.050}& 0.077\scriptsize{$\pm$0.021}\\  
\texttt{\#Monkeypox} 
& \cellcolor[HTML]{EAF3FD}\textbf{0.677}\scriptsize{$\pm$0.020}& 0.624\scriptsize{$\pm$0.071}& 0.045\scriptsize{$\pm$0.026} & \cellcolor[HTML]{EAF3FD}\textbf{0.697}\scriptsize{$\pm$0.014}& 0.656\scriptsize{$\pm$0.045}& 0.057\scriptsize{$\pm$0.011}\\  
\texttt{\#Climatechange} 
& \cellcolor[HTML]{EAF3FD}\textbf{0.664}\scriptsize{$\pm$0.049}& 0.664\scriptsize{$\pm$0.061}& 0.073\scriptsize{$\pm$0.045} & \cellcolor[HTML]{EAF3FD}\textbf{0.715}\scriptsize{$\pm$0.043}& 0.710\scriptsize{$\pm$0.044}& 0.072\scriptsize{$\pm$0.018}\\  
\texttt{\#Brexit} 
& 0.628\scriptsize{$\pm$0.018}& \cellcolor[HTML]{EAF3FD}\textbf{0.685}\scriptsize{$\pm$0.062}& 0.134\scriptsize{$\pm$0.102} & 0.717\scriptsize{$\pm$0.019}& \cellcolor[HTML]{EAF3FD}\textbf{0.744}\scriptsize{$\pm$0.014}& 0.084\scriptsize{$\pm$0.021}\\ 
\texttt{\#IOS16} 
& 0.667\scriptsize{$\pm$0.048}& \cellcolor[HTML]{EAF3FD}\textbf{0.724}\scriptsize{$\pm$0.019}& 0.041\scriptsize{$\pm$0.038} & 0.708\scriptsize{$\pm$0.049}& \cellcolor[HTML]{EAF3FD}\textbf{0.749}\scriptsize{$\pm$0.012}& 0.039\scriptsize{$\pm$0.023}\\  
\texttt{\#Supercup} 
& \cellcolor[HTML]{EAF3FD}\textbf{0.671}\scriptsize{$\pm$0.030}& 0.655\scriptsize{$\pm$0.034}& 0.042\scriptsize{$\pm$0.045} & \cellcolor[HTML]{EAF3FD}\textbf{0.733}\scriptsize{$\pm$0.016}& 0.717\scriptsize{$\pm$0.053}& 0.115\scriptsize{$\pm$0.077}\\  
\texttt{\#Olivianewtonjohn} 
& 0.567\scriptsize{$\pm$0.086}& \cellcolor[HTML]{EAF3FD}\textbf{0.694}\scriptsize{$\pm$0.040}& 0.067\scriptsize{$\pm$0.044} & 0.706\scriptsize{$\pm$0.029}& \cellcolor[HTML]{EAF3FD}\textbf{0.717}\scriptsize{$\pm$0.020}& 0.085\scriptsize{$\pm$0.061}\\  
\texttt{\#CostofLivingCrisis} 
& 0.619\scriptsize{$\pm$0.028}& \cellcolor[HTML]{EAF3FD}\textbf{0.640}\scriptsize{$\pm$0.014}& 0.044\scriptsize{$\pm$0.040} & \cellcolor[HTML]{EAF3FD}\textbf{0.678}\scriptsize{$\pm$0.008}& 0.664\scriptsize{$\pm$0.021}& 0.092\scriptsize{$\pm$0.068}\\  
\texttt{\#Avengers} 
& \cellcolor[HTML]{EAF3FD}\textbf{0.574}\scriptsize{$\pm$0.026}& 0.555\scriptsize{$\pm$0.021}& 0.051\scriptsize{$\pm$0.032} & 0.590\scriptsize{$\pm$0.022}& \cellcolor[HTML]{EAF3FD}\textbf{0.609}\scriptsize{$\pm$0.014}& 0.104\scriptsize{$\pm$0.047}\\  
\texttt{\#UkraineWar} 
& 0.559\scriptsize{$\pm$0.050}& \cellcolor[HTML]{EAF3FD}\textbf{0.588}\scriptsize{$\pm$0.040}& 0.140\scriptsize{$\pm$0.130} & \cellcolor[HTML]{EAF3FD}\textbf{0.567}\scriptsize{$\pm$0.016}& 0.548\scriptsize{$\pm$0.050}& 0.054\scriptsize{$\pm$0.038}\\  
\texttt{\#EnergyPrices} 
& 0.559\scriptsize{$\pm$0.028}& \cellcolor[HTML]{EAF3FD}\textbf{0.576}\scriptsize{$\pm$0.018}& 0.125\scriptsize{$\pm$0.060} & 0.567\scriptsize{$\pm$0.010}& \cellcolor[HTML]{EAF3FD}\textbf{0.575}\scriptsize{$\pm$0.012}& 0.128\scriptsize{$\pm$0.078}\\  
\texttt{\#LoveIsland} 
& 0.501\scriptsize{$\pm$0.059}& \cellcolor[HTML]{EAF3FD}\textbf{0.575}\scriptsize{$\pm$0.013}& 0.138\scriptsize{$\pm$0.057} & 0.569\scriptsize{$\pm$0.016}& \cellcolor[HTML]{EAF3FD}\textbf{0.572}\scriptsize{$\pm$0.023}& 0.151\scriptsize{$\pm$0.115}\\  
\hline
\multirow{2}{*}{ } & \multicolumn{6}{c}{\textbf{Overall} F1 Score ($\bar\mu \pm \bar\sigma$)} \\
& 0.630\scriptsize{$\pm$0.088} & \cellcolor[HTML]{EAF3FD}\textbf{0.649}\scriptsize{$\pm$0.082} & 0.091\scriptsize{$\pm$0.101}& 0.676\scriptsize{$\pm$0.085} & \cellcolor[HTML]{EAF3FD}\textbf{0.687}\scriptsize{$\pm$0.088} & 0.100\scriptsize{$\pm$0.100} \\  
\bottomrule

\multicolumn{7}{p{0.93\linewidth}}{\footnotesize
We {\setlength{\fboxsep}{1pt}\colorbox[HTML]{EAF3FD}{\textbf{highlight}}} the best performance among three feature settings for each hashtag and each model.}\\
\multicolumn{7}{p{0.93\linewidth}}{\footnotesize Note: 
The above results are based on data with the 1:5 class ratio, where F1 score of random guessing is 0.167; and F1 score is 0.286 (or 0.000) if all instances are predicted as positive (or negative). The F1 score is 0.250 if half of the instances are predicted as positive. Further details are provided in Appendix~\ref{app:sec_random_guess}.
We also provide values of precision and recall for the DT model in Appendix~\ref{app:sec_gen_DT_other_metrics}. 
}
\end{tabular}

\label{tab:table_4}
\end{table}

\textit{\textbf{Experimental Setting}.}
We conduct out-of-distribution (OOD) prediction for each of the 14 hashtags. 
Specifically, for each given (held-out) hashtag, we train a model using data from the other 13 hashtags and test it on the held-out hashtag, which represents the new unseen topic. 
This split creates an out-of-distribution condition because every test instance comes from a hashtag absent during training.
This evaluation is important for studying reposting behaviour in dynamic information environments, where social media topics often emerge after a model has already been trained. It therefore tests whether reposting prediction models can generalise from known topics to a previously unseen topic, rather than only predicting held-out instances from the same topics.

To obtain a stable and reliable estimate of model performance, for each given hashtag, we randomly divided the training data of the other 13 hashtags into 10 subsets. 
In each run, one subset is used for validation, and the remaining 9 subsets are used for training.
We repeated this process 10 times by leaving out a different validation subset each time. The 10 trained models are evaluated against the same test data of the given held-out hashtag, producing the mean and standard deviation of F1 scores.
Finally, we aggregate results across the 14 held-out hashtags using the same procedure as in per-hashtag prediction.

\textit{\textbf{Results}.}
Table \ref{tab:table_4} shows the out-of-distribution prediction results. 
When based only on post-related features, all models show notably poor performance. Their overall F1 scores, $\bar\mu$, across 14 hashtags are as low as 0.117, 0.083, 0.091 and 0.100, respectively, which are close to the F1 score of random guessing (see notes in Table~\ref{tab:table_4}).  
The low performance of the M-only models is unlikely to be explained solely by model architecture. Rather, it reflects a limitation of relying exclusively on post text. These M-only models learn from the textual content of posts and therefore are bounded by the vocabularies, entities, and semantic contexts present in the training topics. This constrains their ability to generalise to new topics that are absent from the training data. 

By comparison, models using ALL or U perform substantially better.
Remarkably, models using only user-related features produce performance as good as, or even better than, the corresponding models using all features.
This suggests that user-related features are crucial for out-of-distribution prediction, where adding post-related features would have minimal benefit, or even negative impact, on performance.
More specifically, user-related features provide more transferable predictive signals than post-related features for unseen topics.

Both paired t-tests and Wilcoxon signed-rank tests confirm that the above performance rankings are statistically significant (with all values of \textit{p} $<$ 0.05).
We also provide precision and recall for the DT model in Appendix~\ref{app:sec_gen_DT_other_metrics}. The model using only post-related features consistently exhibits substantially lower precision and recall than the model using user-related features.
We further evaluate inverted predictions for DT-M, where all predicted labels are flipped, to examine whether the poor performance of the M-only model is caused by predictions that systematically favour the opposite class.
Even after inversion, the overall F1 score (0.278) remains low relative to the U and ALL settings.
Further details are provided in Appendix~\ref{app:sec_inverted_pred}.

\section{Discussion}

\subsection{In-Distribution Predictions vs. Out-of-Distribution Prediction}

\begin{figure}[!t]
\centering
\includegraphics[width=0.8\textwidth]{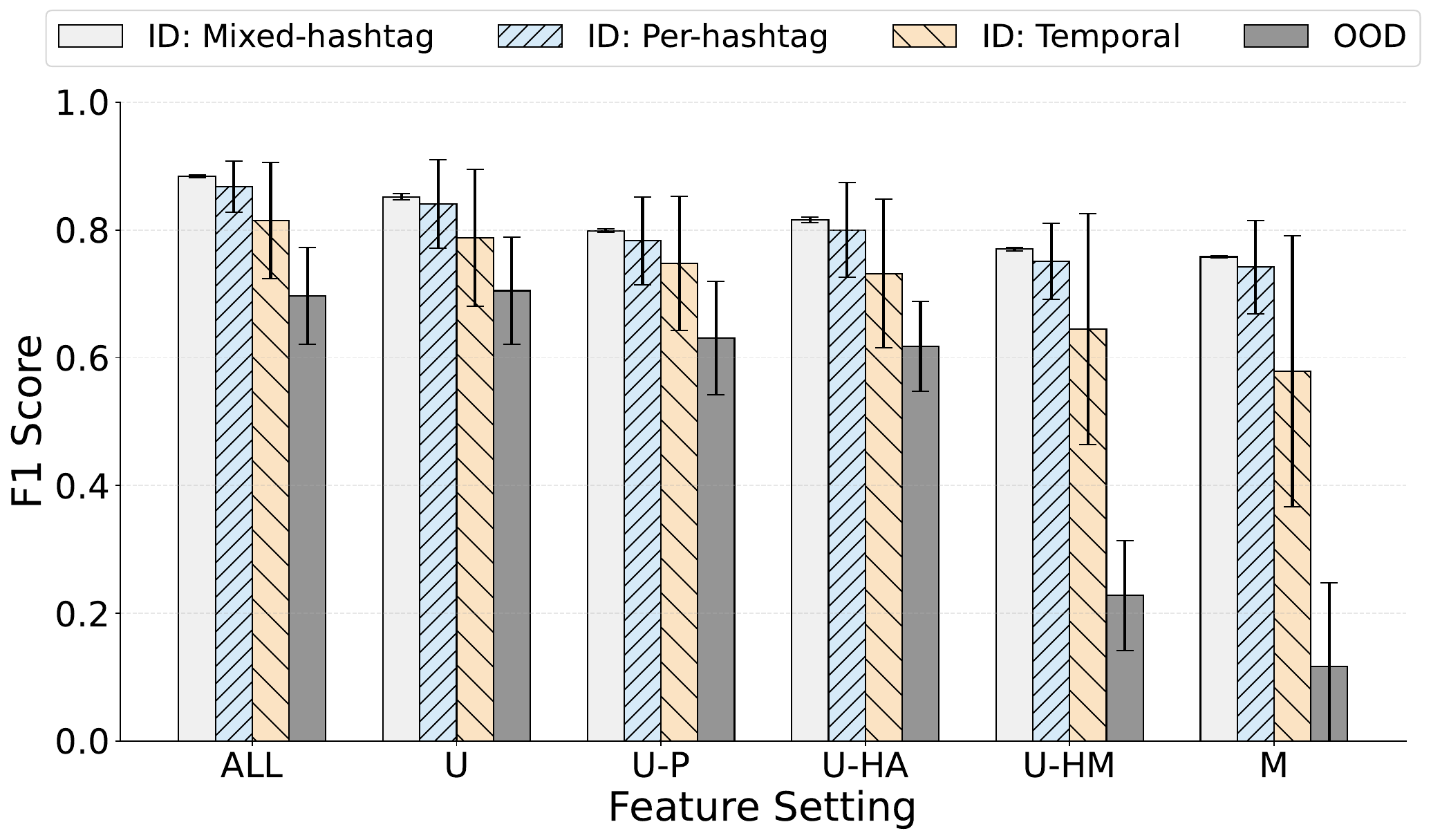}
\caption{
\textbf{In-Distribution (ID)} predictions vs \textbf{Out-Of-Distribution (OOD)} prediction using the Decision Tree model with different feature settings.
\textbf{ID: Mixed-hashtag} is the in-distribution prediction using random splits over mixed hashtags; \textbf{ID: Per-hashtag} is the prediction using random splits within each hashtag; \textbf{ID: Temporal} is the temporal prediction using earlier time windows to predict later ones within each hashtag; and \textbf{OOD} is the out-of-distribution prediction using other hashtags to predict an unseen hashtag. Six different feature settings are studied, where the Decision Tree model uses different types of features: \textbf{ALL} (all features), \textbf{U} (all user-related features including U-P, U-HA, and U-HM), \textbf{U-P} (user profile features), \textbf{U-HA} (user historical action features), \textbf{U-HM} (user historical message features), and \textbf{M} (post-related message features).} 
\label{fig:figure_4} 
\end{figure}

Figure \ref{fig:figure_4} shows that the decision tree model performs worse under out-of-distribution prediction than under the three in-distribution settings: mixed-hashtag, per-hashtag, and temporal prediction.
This drop is expected because out-of-distribution prediction tests the model on a topic absent from training, which is more realistic and challenging, whereas in-distribution prediction uses overlapping topic distributions for training and testing.  

The gap between the two predictions is particularly pronounced for the model using message features M only, whose performance drops substantially for out-of-distribution prediction. 
This indicates that M features can support prediction when training and test data share the same topic distribution, but they are less transferable when the test topic is unseen.
By contrast, user-related features retain much stronger predictive performance under topic shift, suggesting that reposting decisions are largely content-agnostic: they are driven more by stable user characteristics than by the specific content of a post.

The model using U-HM features, extracted from the message text of users' historical posts, behaves more like the M-only model than like the U-P and U-HA models.
Its out-of-distribution performance declines noticeably, though not as severely as M. 
This suggests that historical message text is also sensitive to topic shift.
It contributes less to out-of-distribution prediction than user profile features (U-P) and user historical action features (U-HA), which capture more stable cross-topic signals such as user status, activity patterns, and interaction tendencies.

\subsection{Model Comparison vs. Feature Comparison}
 
The DT model achieves the highest overall performance in our experiments.
One possible explanation is that decision trees can disregard uninformative features through their feature-splitting mechanism: at each node, the tree chooses the feature that best separates the classes, which not only reveals useful signals hidden in the data, but also effectively ignores weak or irrelevant features. 
This result is consistent with the broader observation that tree-based models remain strong baselines when sufficient labelled data and well-engineered features are available~\cite{hegselmann2023tabllm}.

Although DT achieves the highest overall performance, the main observation is not a model ranking.
Rather, the relative advantage of user-related features over post-related features remains consistent across model families.
This is consistent with the broader view that cognitive and behavioural mechanisms are not peripheral details but fundamental determinants of information diffusion~\cite{caldarelli2025physics}.

\subsection{Implications for Modelling Information Diffusion}

Our findings have two broader implications for modelling information diffusion.
First, they suggest that evaluating information diffusion models solely through in-distribution experiments can provide an incomplete picture of their practical usefulness.
In-distribution experiments provide a useful baseline for measuring performance when future instances resemble past instances.
The out-of-distribution experiment therefore complements them by testing whether models remain effective under distribution shift.

Second, our findings suggest that reposting prediction may depend less on understanding the exact semantic meaning of posts than on modelling the behavioural and social characteristics of users interacting with those posts.
This does not mean that post content is irrelevant: post-related features remain useful in in-distribution settings.
Rather, the results show that post-related and user-related signals generalise differently under topic shift, and that improving prediction for unseen topics may require stronger user modelling and behavioural representation rather than relying solely on increasingly powerful text representations.

\subsection{Future Work}

A recent work based on LLM-agent simulations ~\cite{larooij2025can} shows that users can still form polarised echo chambers on social networks even without content-based recommendation algorithms.
This points to an important direction for future work: studying how user characteristics, post content, and algorithmic exposure jointly influence information diffusion.

In addition, our results provide predictive evidence that user-related features are important for micro-level reposting decisions. A natural next step is to move beyond predictive modelling and examine causal mechanisms and platform-level interventions, including how changes in exposure, recommendation algorithms, or social context may affect reposting decisions and broader diffusion outcomes.

Text remains the primary modality used in prior reposting prediction research. As multimodal content becomes increasingly common on social media, future work should also study how images and videos embedded in posts affect reposting decisions.

\section{Conclusion}

Reposting prediction is a fundamental problem in modelling information diffusion.
In this work, we studied micro-level reposting decisions that form the basis of macro-level information diffusion. Specifically, we investigated the task of predicting whether a user will repost a specific post from another user.
We applied three in-distribution evaluation settings and a cross-topic out-of-distribution evaluation setting in which models are trained on known topics and tested on a previously unseen topic. We systematically compared post-related features, user-related features, and their combination across four model families and four evaluation setups.
The results revealed a consistent pattern: while post content can be useful when training and test topics are similar, user characteristics play a more important role in reposting prediction and provide transferable signals for unseen topics. 
These findings highlight the importance of user modelling when seeking to understand and predict information diffusion in online social networks.

\section*{Acknowledgments} 
I.J.C. and V.L. would like to acknowledge support from the EPSRC grant EP/X031276/1. The authors thank Ziwen Li, Berkem Billuroglu, and Pei Lo for their contribution to the X data collection and preliminary analysis during their study at UCL.

\bibliographystyle{ACM-Reference-Format}
\bibliography{reference}

@String{Computing = "Computing" }

@String{Computer = "{IEEE} Computer" }

@String{Springer = "Springer-Verlag" }

@article{zhu2023path,
  title={{Path prediction of information diffusion based on a topic-oriented relationship strength network}},
  author={Zhu, Hengmin and Yang, Xinyi and Wei, Jing},
  journal={Information Sciences},
  volume={631},
  pages={108--119},
  year={2023},
  publisher={Elsevier}
}

@inproceedings{antypas2022twitter,
  title={{Twitter Topic Classification}},
  author={Antypas, Dimosthenis and Ushio, Asahi and Camacho-Collados, Jose and Silva, V{\'\i}tor and Neves, Leonardo and Barbieri, Francesco},
  booktitle={Proceedings of the 29th International Conference on Computational Linguistics},
  pages={3386--3400},
  year={2022},
  address={Gyeongju, Republic of Korea},
  publisher={International Committee on Computational Linguistics},
}

@inproceedings{hong2011predicting,
  title={{Predicting popular messages in Twitter}},
  author={Hong, Liangjie and Dan, Ovidiu and Davison, Brian D},
  booktitle={Proceedings of the 20th International Conference Companion on World Wide Web},
  pages={57--58},
  year={2011},
  publisher={Association for Computing Machinery},
  address={New York, NY, USA}
}

@inproceedings{martin2016exploring,
  title={{Exploring limits to prediction in complex social systems}},
  author={Martin, Travis and Hofman, Jake M and Sharma, Amit and Anderson, Ashton and Watts, Duncan J},
  booktitle={Proceedings of the 25th International Conference on World Wide Web},
  pages={683--694},
  year={2016},
  publisher={International World Wide Web Conferences Steering Committee},
  address={Republic and Canton of Geneva, CHE}
}

@inproceedings{barbieri2020tweeteval,
  title={{TweetEval: Unified benchmark and comparative evaluation for tweet classification}},
  author={Barbieri, Francesco and Camacho-Collados, Jose and Anke, Luis Espinosa and Neves, Leonardo},
  booktitle={Findings of the Association for Computational Linguistics: EMNLP 2020},
  pages={1644--1650},
  year={2020},
  address={Online},
  publisher={Association for Computational Linguistics}
}

@article{hutto2014vader,
  title={{Vader: A parsimonious rule-based model for sentiment analysis of social media text}},
  author={Hutto, Clayton and Gilbert, Eric},
  journal={Proceedings of the international AAAI conference on web and social media},
  volume={8},
  pages={216--225},
  year={2014}
}

@article{khabiri2009analyzing,
  title={{Analyzing and predicting community preference of socially generated metadata: A case study on comments in the digg community}},
  author={Khabiri, Elham and Hsu, Chiao-Fang and Caverlee, James},
  journal={Proceedings of the International AAAI Conference on Web and Social Media},
  volume={3},
  pages={238--241},
  year={2009}
}

@article{robertson2023negativity,
  title={{Negativity drives online news consumption}},
  author={Robertson, Claire E and Pr{\"o}llochs, Nicolas and Schwarzenegger, Kaoru and P{\"a}rnamets, Philip and Van Bavel, Jay J and Feuerriegel, Stefan},
  journal={Nature human behaviour},
  volume={7},
  number={5},
  pages={812--822},
  year={2023},
  publisher={Nature Publishing Group UK London}
}

@inproceedings{zhao2018comparative,
  title={{A comparative study of transactional and semantic approaches for predicting cascades on Twitter}},
  author={Zhao, Yunwei and Wang, Can and Chi, Chi-Hung and Lam, Kwok-Yan and Wang, Sen},
  booktitle = {Proceedings of the 27th International Joint Conference on Artificial Intelligence},
  publisher={AAAI Press},
  pages={1212--1218},
  year={2018},
  month={7},
  doi={10.24963/ijcai.2018/169},
  url={https://doi.org/10.24963/ijcai.2018/169}
}

@article{paletz2023emotional,
  title={{Emotional content and sharing on Facebook: A theory cage match}},
  author={Paletz, Susannah BF and Johns, Michael A and Murauskaite, Egle E and Golonka, Ewa M and Pand{\v{z}}a, Nick B and Rytting, C Anton and Buntain, Cody and Ellis, Devin},
  journal={Science Advances},
  volume={9},
  number={39},
  pages={eade9231},
  year={2023},
  publisher={American Association for the Advancement of Sci.}
}

@article{wang2021effect,
  title={{Effect of the lockdown on diurnal patterns of emotion expression in Twitter}},
  author={Wang, Sheng and Lightman, Stafford and Cristianini, Nello},
  journal={Chronobiology International},
  volume={38},
  number={11},
  pages={1591--1610},
  year={2021},
  publisher={Taylor \& Francis}
}

@inproceedings{mousavi2022effective,
  title={{Effective messaging on social media: What makes online content go viral?}},
  author={Mousavi, Maryam and Davulcu, Hasan and Ahmadi, Mohsen and Axelrod, Robert and Davis, Richard and Atran, Scott},
  booktitle={Proceedings of the ACM web conference 2022},
  pages={2957--2966},
  year={2022},
  publisher={Association for Computing Machinery},
  address={New York, NY, USA}
}

@inproceedings{cinus2025exposing,
  title={{Exposing cross-platform coordinated inauthentic activity in the run-up to the 2024 us election}},
  author={Cinus, Federico and Minici, Marco and Luceri, Luca and Ferrara, Emilio},
  booktitle={Proceedings of the ACM web conference 2025},
  pages={541--559},
  year={2025},
  publisher={Association for Computing Machinery},
  address={New York, NY, USA}
}

@inproceedings{nettasinghe2025group,
  title={{In-Group Love, Out-Group Hate: A Framework to Measure Affective Polarization via Contentious Online Discussions}},
  author={Nettasinghe, Buddhika and Rao, Ashwin and Jiang, Bohan and Percus, Allon G and Lerman, Kristina},
  booktitle={Proceedings of the ACM web conference 2025},
  pages={560--575},
  year={2025},
  publisher={Association for Computing Machinery},
  address={New York, NY, USA}
}

@inproceedings{ashkinaze2024dynamics,
  title={{The dynamics of (not) unfollowing misinformation spreaders}},
  author={Ashkinaze, Joshua and Gilbert, Eric and Budak, Ceren},
  booktitle={Proceedings of the ACM web conference 2024},
  pages={1115--1125},
  year={2024},
  publisher={Association for Computing Machinery},
  address={New York, NY, USA}
}

@article{zhou2020realistic,
  title={{Realistic modelling of information spread using peer-to-peer diffusion patterns}},
  author={Zhou, Bin and Pei, Sen and Muchnik, Lev and Meng, Xiangyi and Xu, Xiaoke and Sela, Alon and Havlin, Shlomo and Stanley, H Eugene},
  journal={Nature human behaviour},
  volume={4},
  number={11},
  pages={1198--1207},
  year={2020},
  publisher={Nature Publishing Group UK London}
}

@misc{huberman2008social,
  title={{Social networks that matter: Twitter under the microscope}},
  author={Huberman, Bernardo A and Romero, Daniel M and Wu, Fang},
  year={2008},
  eprint={0812.1045},
  archivePrefix={arXiv},
  primaryClass={cs.CY},
  url={https://arxiv.org/abs/0812.1045}
}

@inproceedings{bakshy2011everyone,
  title={{Everyone's an influencer: quantifying influence on Twitter}},
  author={Bakshy, Eytan and Hofman, Jake M and Mason, Winter A and Watts, Duncan J},
  booktitle={Proceedings of the Fourth ACM International Conference on Web Search and Data Mining},
  pages={65--74},
  year={2011},
  publisher={Association for Computing Machinery},
  address={New York, NY, USA}
}

@inproceedings{islam2018deepdiffuse,
  title={{Deepdiffuse: Predicting the 'who' and 'when' in cascades}},
  author={Islam, Mohammad Raihanul and Muthiah, Sathappan and Adhikari, Bijaya and Prakash, B Aditya and Ramakrishnan, Naren},
  booktitle={2018 IEEE International Conference on Data Mining (ICDM)},
  pages={1055--1060},
  year={2018},
  publisher={{IEEE} Computer Society},
  
}

@inproceedings{yuan2021dyhgcn,
  title={{DyHGCN: A dynamic heterogeneous graph convolutional network to learn users’ dynamic preferences for information diffusion prediction}},
  author={Yuan, Chunyuan and Li, Jiacheng and Zhou, Wei and Lu, Yijun and Zhang, Xiaodan and Hu, Songlin},
  booktitle={Joint European Conference on Machine Learning and Knowledge Discovery in Databases},
  pages={347--363},
  year={2020},
  publisher = {Springer-Verlag},
  address = {Berlin, Heidelberg}
}

@inproceedings{zhang2016retweet,
  title={{Retweet prediction with attention-based deep neural network}},
  author={Zhang, Qi and Gong, Yeyun and Wu, Jindou and Huang, Haoran and Huang, Xuanjing},
  booktitle={Proceedings of the 25th ACM International on Conference on Information and Knowledge Management},
  pages={75--84},
  year={2016},
  publisher={Association for Computing Machinery},
  address={New York, NY, USA}
}

@inproceedings{feng2013retweet,
  title={{Retweet or not? Personalized tweet re-ranking}},
  author={Feng, Wei and Wang, Jianyong},
  booktitle={Proceedings of the Sixth ACM International Conference on Web Search and Data Mining},
  pages={577--586},
  year={2013},
  publisher={Association for Computing Machinery},
  address={New York, NY, USA}
}

@inproceedings{luo2013will,
  title={{Who will retweet me? Finding retweeters in Twitter}},
  author={Luo, Zhunchen and Osborne, Miles and Tang, Jintao and Wang, Ting},
  booktitle={Proceedings of the 36th International ACM SIGIR conference on Research and Development in Information Retrieval},
  pages={869--872},
  year={2013},
  publisher={Association for Computing Machinery},
  address={New York, NY, USA}
}

@inproceedings{cao2021information,
  title={{Information diffusion prediction via dynamic graph neural networks}},
  author={Cao, Zongmai and Han, Kai and Zhu, Jianfu},
  booktitle={2021 IEEE 24th International Conference on Computer Supported Cooperative Work in Design (CSCWD)},
  pages={1099--1104},
  year={2021},
  publisher={IEEE}
}

@inproceedings{xu2012analyzing,
  title={{Analyzing user retweet behavior on Twitter}},
  author={Xu, Zhiheng and Yang, Qing},
  booktitle={2012 IEEE/ACM International Conference on Advances in Social Networks Analysis and Mining},
  pages={46--50},
  year={2012},
  publisher={{IEEE} Computer Society},
  address={USA}
}

@article{yang2021full,
  title={{Full-scale information diffusion prediction with reinforced recurrent networks}},
  author={Yang, Cheng and Wang, Hao and Tang, Jian and Shi, Chuan and Sun, Maosong and Cui, Ganqu and Liu, Zhiyuan},
  journal={IEEE Transactions on Neural Networks and Learning Systems},
  volume={34},
  number={5},
  pages={2271--2283},
  year={2021},
  publisher={IEEE}
}

@inproceedings{ma2019hot,
  title={{Hot topic-aware retweet prediction with masked self-attentive model}},
  author={Ma, Renfeng and Hu, Xiangkun and Zhang, Qi and Huang, Xuanjing and Jiang, Yu-Gang},
  booktitle={Proceedings of the 42nd international ACM SIGIR conference on research and development in information retrieval},
  pages={525--534},
  year={2019},
  publisher={ACM}
}

@article{firdaus2021retweet,
  title={{Retweet prediction based on topic, emotion and personality}},
  author={Firdaus, Syeda Nadia and Ding, Chen and Sadeghian, Alireza},
  journal={Online Social Networks and Media},
  volume={25},
  pages={100165},
  year={2021},
  publisher={Elsevier}
}

@inproceedings{jiang2018retweet,
  title={{Retweet prediction using social-aware probabilistic matrix factorization}},
  author={Jiang, Bo and Lu, Zhigang and Li, Ning and Wu, Jianjun and Jiang, Zhengwei},
  booktitle={Computational Science – ICCS 2018: 18th International Conference, Wuxi, China, June 11–13, 2018, Proceedings, Part I},
  pages={316--327},
  year={2018},
  publisher = {Springer-Verlag},
  address = {Berlin, Heidelberg}
}

@inproceedings{jiang2019retweeting,
  title={{Retweeting prediction using matrix factorization with binomial distribution and contextual information}},
  author={Jiang, Bo and Lu, Zhigang and Li, Ning and Wu, Jianjun and Yi, Feng and Han, Dongxu},
  booktitle={International Conference on Database Systems for Advanced Applications},
  pages={121--138},
  year={2019},
  publisher={Springer International Publishing},
  address={Cham}
}

@inproceedings{wang2017topological,
  title={{Topological recurrent neural network for diffusion prediction}},
  author={Wang, Jia and Zheng, Vincent W and Liu, Zemin and Chang, Kevin Chen-Chuan},
  booktitle={2017 IEEE international conference on data mining (ICDM)},
  pages={475--484},
  year={2017},
  publisher={{IEEE} Computer Society}
}

@article{quan2018repost,
  title={{Repost prediction incorporating time-sensitive mutual influence in social networks}},
  author={Quan, Yong and Jia, Yan and Zhou, Bin and Han, Weihong and Li, Shudong},
  journal={Journal of Computational Science},
  volume={28},
  pages={217--227},
  year={2018},
  publisher={Elsevier}
}

@inproceedings{yang2017propagator,
  title={{Propagator or influencer? A data-driven approach for evaluating emotional effect in online information diffusion}},
  author={Yang, Jun and Wang, Zhaoguo and Di, Fangchun and Chen, Liyue and Yi, Chengqi and Xue, Yibo and Li, Jun},
  booktitle={Proceedings of the 2017 IEEE/ACM International Conference on Advances in Social Networks Analysis and Mining 2017},
  pages={836--843},
  year={2017},
  publisher = {Association for Computing Machinery},
  address = {New York, NY, USA}
}

@inproceedings{zhang2013social,
  title={{Social influence locality for modeling retweeting behaviors}},
  author={Zhang, Jing and Liu, Biao and Tang, Jie and Chen, Ting and Li, Juanzi},
  booktitle={Proceedings of the Twenty-Third International Joint Conference on Artificial Intelligence},
  publisher={AAAI Press},
  volume={13},
  pages={2761--2767},
  year={2013}
}

@inproceedings{tu2022viral,
  title={{A viral marketing-based model for opinion dynamics in online social networks}},
  author={Tu, Sijing and Neumann, Stefan},
  booktitle={Proceedings of the ACM web conference 2022},
  pages={1570--1578},
  year={2022},
  publisher={Association for Computing Machinery},
  address={New York, NY, USA}
}

@ARTICLE{chen2020event,
  title={{Event popularity prediction using influential hashtags from social media}},
  author={Chen, Xi and Zhou, Xiangmin and Chan, Jeffrey and Chen, Lei and Sellis, Timos and Zhang, Yanchun},
  journal={IEEE Transactions on Knowledge and Data Engineering},  
  volume={34},
  number={10},
  pages={4797-4811},
  year={2022},
  publisher={IEEE}
}

@article{hoang2016microblogging,
  title={{Microblogging content propagation modeling using topic-specific behavioral factors}},
  author={Hoang, Tuan-Anh and Lim, Ee-Peng},
  journal={IEEE Transactions on Knowledge and Data Engineering},
  volume={28},
  number={9},
  pages={2407--2422},
  year={2016},
  publisher={IEEE}
}

@ARTICLE{xu2021casflow,
  title={{CasFlow: Exploring hierarchical structures and propagation uncertainty for cascade prediction}}, 
  author={Xu, Xovee and Zhou, Fan and Zhang, Kunpeng and Liu, Siyuan and Trajcevski, Goce},
  journal={IEEE Transactions on Knowledge and Data Engineering}, 
  volume={35},
  number={4},
  pages={3484-3499},
  year={2023},
  publisher={IEEE}
}

@article{yin2021deep,
  title={{Deep fusion of multimodal features for social media retweet time prediction}},
  author={Yin, Hui and Yang, Shuiqiao and Song, Xiangyu and Liu, Wei and Li, Jianxin},
  journal={World Wide Web},
  volume={24},
  number={4},
  pages={1027--1044},
  year={2021},
  publisher={Springer}
}

@article{wang2022tweet,
  title={{Tweet retweet prediction based on deep multitask learning}},
  author={Wang, Jing and Yang, Yue},
  journal={Neural Processing Letters},
  volume={54},
  number={1},
  pages={523--536},
  year={2022},
  publisher={Springer}
}

@inproceedings{li2017deepcas,
  title={{Deepcas: An end-to-end predictor of information cascades}},
  author={Li, Cheng and Ma, Jiaqi and Guo, Xiaoxiao and Mei, Qiaozhu},
  booktitle={Proceedings of the 26th international conference on World Wide Web},
  pages={577--586},
  year={2017},
  publisher = {International World Wide Web Conferences Steering Committee},
  address = {Republic and Canton of Geneva, CHE}
}

@phdthesis{arjovsky2020out,
  title={{Out of distribution generalization in machine learning}},
  author={Arjovsky, Martin},
  year={2020},
  school={New York University}
}

@misc{liu2021towards,
  title={{Towards out-of-distribution generalization: A survey}},
  author={Liu, Jiashuo and Shen, Zheyan and He, Yue and Zhang, Xingxuan and Xu, Renzhe and Yu, Han and Cui, Peng},
  year={2023},
  eprint={2108.13624},
  archivePrefix={arXiv},
  primaryClass={cs.LG},
  url={https://arxiv.org/abs/2108.13624}
}

@article{he2021cannot,
  title={{Cannot predict comment volume of a news article before (a few) users read it}},
  author={He, Lihong and Shen, Chen and Mukherjee, Arjun and Vucetic, Slobodan and Dragut, Eduard},
  journal={Proceedings of the International AAAI Conference on Web and Social Media},
  volume={15},
  pages={173--184},
  year={2021}
}

@article{pfeffer2023half,
  title={{The half-life of a tweet}},
  author={Pfeffer, J{\"u}rgen and Matter, Daniel and Sargsyan, Anahit},
  journal={Proceedings of the International AAAI Conference on Web and Social Media},
  volume={17},
  pages={1163--1167},
  year={2023}
}

@inproceedings{devlin2019bert,
  title={{Bert: Pre-training of deep bidirectional transformers for language understanding}},
  author={Devlin, Jacob and Chang, Ming-Wei and Lee, Kenton and Toutanova, Kristina},
  booktitle={Proceedings of the 2019 Conference of the North American Chapter of the Association for Computational Linguistics: Human Language Technologies, Volume 1 (Long and Short Papers)},
  pages={4171--4186},
  year={2019},
  publisher={Association for Computational Linguistics},
  address={Minneapolis, Minnesota}
}

@inproceedings{lampos2014predicting,
  title={{Predicting and characterising user impact on Twitter}},
  author={Lampos, Vasileios and Aletras, Nikolaos and Preo{\c{t}}iuc-Pietro, Daniel and Cohn, Trevor},
  booktitle={Proceedings of the 14th Conference of the European Chapter of the Association for Computational Linguistics},
  pages={405--413},
  year={2014},
  publisher={The Association for Computer Linguistics}
}

@article{Ma2013OnTwitter,
    title={{On predicting the popularity of newly emerging hashtags in Twitter}},
    author={Ma, Zongyang and Sun, Aixin and Cong, Gao},
    journal={Journal of the American Society for Information Science and Technology},
    volume={64},
    number={7},
    pages={1399--1410},
    year={2013},
    publisher={Wiley Online Library}
}

@article{Zheng2022PredictingNetworks,
    title={{Predicting hot events in the early period through bayesian model for social networks}},
    author={Zheng, Zuowu and Gao, Xiaofeng and Ma, Xiao and Chen, Guihai},
    journal={IEEE Transactions on Knowledge and Data Engineering},
    volume={34},
    number={3},
    pages={1390--1403},
    year={2022},
    publisher={IEEE}
}

@misc{larooij2025can,
  title={{Can We Fix Social Media? Testing Prosocial Interventions using Generative Social Simulation}},
  author={Larooij, Maik and T{\"o}rnberg, Petter},
  year={2025},
  eprint={2508.03385},
  archivePrefix={arXiv},
  primaryClass={cs.SI},
  url={https://arxiv.org/abs/2508.03385}
}

@article{engel2025social,
  title={{Social Dynamics and Mobilization Potential of Online Election Narratives}},
  author={Engel, Kristen and Mitra, Tanushree and Spiro, Emma S},
  journal={Proceedings of the International AAAI Conference on Web and Social Media},
  volume={19},
  pages={479--496},
  year={2025}
}

@article{sakib2025opposites,
  title={{Opposites Attract? Ambivalence in Distinguishing Real and Fake News and Predicting their Spread}},
  author={Sakib, Mostofa Najmus and Spezzano, Francesca and Hamby, Anne},
  journal={Proceedings of the International AAAI Conference on Web and Social Media},
  volume={19},
  pages={2650--2659},
  year={2025}
}

@inproceedings{ergu2019predicting,
  title={{Predicting personality with twitter data and machine learning models}},
  author={Ergu, Izel and I{\c{s}}{\i}k, Zerrin and Yankay{\i}{\c{s}}, {\.I}smail},
  booktitle={2019 innovations in intelligent systems and applications conference (ASYU)},
  pages={1--5},
  year={2019},
  publisher={IEEE}
}

@article{arnoux201725,
  title={{25 tweets to know you: A new model to predict personality with social media}},
  author={Arnoux, Pierre-Hadrien and Xu, Anbang and Boyette, Neil and Mahmud, Jalal and Akkiraju, Rama and Sinha, Vibha},
  journal={Proceedings of the International AAAI Conference on Web and Social Media},
  volume={11(1)},
  pages={472--475},
  year={2017}
}

@article{caldarelli2025physics,
  title={{The physics of news, rumors, and opinions}},
  journal = {Physics Reports},
  volume = {1186},
  pages = {1-75},
  year = {2026},
  issn = {0370-1573},
  doi = {https://doi.org/10.1016/j.physrep.2026.05.002},
  author = {Guido Caldarelli and Oriol Artime and Giulia Fischetti and Stefano Guarino and Andrzej Nowak and Fabio Saracco and Petter Holme and Manlio {De Domenico}},
}

@article{dormann2013collinearity,
  title={{Collinearity: a review of methods to deal with it and a simulation study evaluating their performance}},
  author={Dormann, Carsten F and Elith, Jane and Bacher, Sven and Buchmann, Carsten and Carl, Gudrun and Carr{\'e}, Gabriel and Marqu{\'e}z, Jaime R Garc{\'\i}a and Gruber, Bernd and Lafourcade, Bruno and Leit{\~a}o, Pedro J and others},
  journal={Ecography},
  volume={36},
  number={1},
  pages={27--46},
  year={2013},
  publisher={Wiley Online Library}
}

@article{zhang2019deep,
  title={{Deep learning based recommender system: A survey and new perspectives}},
  author={Zhang, Shuai and Yao, Lina and Sun, Aixin and Tay, Yi},
  journal={ACM computing surveys (CSUR)},
  volume={52},
  number={1},
  pages={1--38},
  year={2019},
  publisher={ACM New York, NY, USA}
}

@article{wu2022graph,
  title={{Graph neural networks in recommender systems: a survey}},
  author={Wu, Shiwen and Sun, Fei and Zhang, Wentao and Xie, Xu and Cui, Bin},
  journal={ACM Computing Surveys},
  volume={55},
  number={5},
  pages={1--37},
  year={2022},
  publisher={ACM New York, NY}
}

@article{bobadilla2013recommender,
  title={{Recommender systems survey}},
  author={Bobadilla, Jes{\'u}s and Ortega, Fernando and Hernando, Antonio and Guti{\'e}rrez, Abraham},
  journal={Knowledge-based systems},
  volume={46},
  pages={109--132},
  year={2013},
  publisher={Elsevier}
}

@inproceedings{yan2012tweet,
  title={{Tweet recommendation with graph co-ranking}},
  author={Yan, Rui and Lapata, Mirella and Li, Xiaoming},
  booktitle={The 50th Annual Meeting of the Association for Computational Linguistics, Proceedings of the Conference, July 8-14, 2012, Jeju Island, Korea-Volume 1: Long Papers},
  pages={516--525},
  year={2012},
  publisher={Association for Computational Linguistics},
  address={USA}
}

@inproceedings{diaz2014predicting,
  title={{Predicting user engagement in twitter with collaborative ranking}},
  author={Diaz-Aviles, Ernesto and Lam, Hoang Thanh and Pinelli, Fabio and Braghin, Stefano and Gkoufas, Yiannis and Berlingerio, Michele and Calabrese, Francesco},
  booktitle={Proceedings of the 2014 Recommender Systems Challenge},
  pages={41--46},
  year={2014},
  publisher={Association for Computing Machinery},
  address={New York, NY, USA}
}

@inproceedings{alawad2016network,
  title={Network-aware recommendations of novel tweets},
  author={Alawad, Noor Aldeen and Anagnostopoulos, Aris and Leonardi, Stefano and Mele, Ida and Silvestri, Fabrizio},
  booktitle={Proceedings of the 39th International ACM SIGIR conference on Research and Development in Information Retrieval},
  pages={913--916},
  year={2016},
  publisher={Association for Computing Machinery},
  address={New York, NY, USA}
}

@article{jerez2026human,
author = {Jerez, Eleana and Jurado, Francisco and Moreno-Llorena, Jaime},
title = {{A Human-Centred Profiling Approach to Characterise User Roles in Information Propagation on Online Social Networks}},
year = {2026},
publisher = {Association for Computing Machinery},
address = {New York, NY, USA},
issn = {1559-1131},
url = {https://doi.org/10.1145/3787206},
doi = {10.1145/3787206},
journal = {ACM Transactions on the Web},
note = {Just Accepted}
}

@article{tang2025msa,
author = {Tang, Yinzhou and Piao, Jinghua and Wang, Huandong and Wang, Yue and Li, Yong},
title = {{MSA-Net: A Multi-Scale Information Diffusion Model Awaring User Activity Level}},
year = {2025},
issue_date = {May 2025},
publisher = {Association for Computing Machinery},
address = {New York, NY, USA},
volume = {19},
number = {2},
issn = {1559-1131},
url = {https://doi.org/10.1145/3711911},
doi = {10.1145/3711911},
journal = {ACM Transactions on the Web},
month = may,
articleno = {17},
numpages = {23},
}

@article{gu2025online,
author = {Gu, Haoran and Zheng, Shiyuan and Liu, Xudong and Xie, Hong and Lui, John C.S.},
title = {{Online Incentive Protocol Design for Reposting Service in Online Social Networks}},
year = {2025},
issue_date = {May 2025},
publisher = {Association for Computing Machinery},
address = {New York, NY, USA},
volume = {19},
number = {2},
issn = {1559-1131},
url = {https://doi.org/10.1145/3696473},
doi = {10.1145/3696473},
journal = {ACM Transactions on the Web},
month = may,
articleno = {20},
numpages = {30}
}

@article{impicciche2025comparing,
author = {Impiccich\`{e}, Paola and Viviani, Marco},
title = {{Comparing Echo Chamber Detection Metrics: A Cross-modeling and Cross-platform Analysis of Twitter and Reddit}},
year = {2025},
issue_date = {November 2025},
publisher = {Association for Computing Machinery},
address = {New York, NY, USA},
volume = {19},
number = {4},
issn = {1559-1131},
url = {https://doi.org/10.1145/3707701},
doi = {10.1145/3707701},
journal = {ACM Transactions on the Web},
month = oct,
articleno = {43},
numpages = {23}
}

@article{evangelatos2025modeling,
author = {Evangelatos, Spyridon and Veroni, Eleni and Efthymiou, Vasilis and Nikolopoulos, Christos},
title = {{Modeling Disinformation Spread in Social Networks: Phase Transitions and Mean-Field Analysis}},
year = {2025},
issue_date = {November 2025},
publisher = {Association for Computing Machinery},
address = {New York, NY, USA},
volume = {19},
number = {4},
issn = {1559-1131},
url = {https://doi.org/10.1145/3747287},
doi = {10.1145/3747287},
journal = {ACM Transactions on the Web},
month = oct,
articleno = {46},
numpages = {24}
}

@article{upadhyaya2022spotting,
author = {Upadhyaya, Apoorva and Chandra, Joydeep},
title = {{Spotting Flares: The Vital Signs of the Viral Spread of Tweets Made During Communal Incidents}},
year = {2022},
issue_date = {November 2022},
publisher = {Association for Computing Machinery},
address = {New York, NY, USA},
volume = {16},
number = {4},
issn = {1559-1131},
url = {https://doi.org/10.1145/3550357},
doi = {10.1145/3550357},
journal = {ACM Transactions on the Web},
month = nov,
articleno = {18},
numpages = {28}
}

@misc{yang2025qwen3technicalreport,
      title={{Qwen3 Technical Report}}, 
      author={An Yang and Anfeng Li and Baosong Yang and Beichen Zhang and Binyuan Hui and Bo Zheng and Bowen Yu and Chang Gao and Chengen Huang and Chenxu Lv and Chujie Zheng and Dayiheng Liu and Fan Zhou and Fei Huang and Feng Hu and Hao Ge and Haoran Wei and Huan Lin and Jialong Tang and Jian Yang and Jianhong Tu and Jianwei Zhang and Jianxin Yang and Jiaxi Yang and Jing Zhou and Jingren Zhou and Junyang Lin and Kai Dang and Keqin Bao and Kexin Yang and Le Yu and Lianghao Deng and Mei Li and Mingfeng Xue and Mingze Li and Pei Zhang and Peng Wang and Qin Zhu and Rui Men and Ruize Gao and Shixuan Liu and Shuang Luo and Tianhao Li and Tianyi Tang and Wenbiao Yin and Xingzhang Ren and Xinyu Wang and Xinyu Zhang and Xuancheng Ren and Yang Fan and Yang Su and Yichang Zhang and Yinger Zhang and Yu Wan and Yuqiong Liu and Zekun Wang and Zeyu Cui and Zhenru Zhang and Zhipeng Zhou and Zihan Qiu},
      year={2025},
      eprint={2505.09388},
      archivePrefix={arXiv},
      primaryClass={cs.CL},
      url={https://arxiv.org/abs/2505.09388}, 
}

@inproceedings{hegselmann2023tabllm,
  title={{Tabllm: Few-shot classification of tabular data with large language models}},
  author={Hegselmann, Stefan and Buendia, Alejandro and Lang, Hunter and Agrawal, Monica and Jiang, Xiaoyi and Sontag, David},
  booktitle={International conference on artificial intelligence and statistics},
  pages={5549--5581},
  year={2023},
  organization={PMLR}
}

@article{hu2022lora,
  title={{Lora: Low-rank adaptation of large language models}},
  author={Hu, Edward J and Shen, Yelong and Wallis, Phillip and Allen-Zhu, Zeyuan and Li, Yuanzhi and Wang, Shean and Wang, Liang and Chen, Weizhu and others},
  journal={ICLR},
  volume={1},
  number={2},
  pages={3},
  year={2022}
}

@inproceedings{zheng-etal-2024-llamafactory,
    title={{LlamaFactory: Unified Efficient Fine-Tuning of 100+ Language Models}},
    author={Zheng, Yaowei and Zhang, Richong and Zhang, Junhao and Ye, Yanhan and Luo, Zheyan},
    booktitle = {Proceedings of the 62nd Annual Meeting of the Association for Computational Linguistics (Volume 3: System Demonstrations)},
    month = aug,
    year = {2024},
    address = {Bangkok, Thailand},
    publisher = {Association for Computational Linguistics},
    url = {https://aclanthology.org/2024.acl-demos.38/},
    doi = {10.18653/v1/2024.acl-demos.38},
    pages = {400--410}
}

@article{jiang2023retweet,
  title={Retweet-bert: political leaning detection using language features and information diffusion on social networks},
  author={Jiang, Julie and Ren, Xiang and Ferrara, Emilio},
  journal={Proceedings of the international AAAI conference on web and social media},
  volume={17},
  pages={459--469},
  year={2023}
}

@article{jiang2025social,
  title={Social-llm: Modeling user behavior at scale using language models and social network data},
  author={Jiang, Julie and Ferrara, Emilio},
  journal={Sci},
  volume={7},
  number={4},
  pages={138},
  year={2025},
  publisher={MDPI}
}

@misc{guo2025somermultiviewuserrepresentation,
      title={SoMeR: Multi-View User Representation Learning for Social Media}, 
      author={Siyi Guo and Keith Burghardt and Valeria Pantè and Kristina Lerman},
      year={2025},
      eprint={2405.05275},
      archivePrefix={arXiv},
      primaryClass={cs.SI},
      url={https://arxiv.org/abs/2405.05275}, 
}

@article{matakos2020tell,
  title={Tell me something my friends do not know: Diversity maximization in social networks},
  author={Matakos, Antonis and Tu, Sijing and Gionis, Aristides},
  journal={Knowledge and Information Systems},
  volume={62},
  number={9},
  pages={3697--3726},
  year={2020},
  publisher={Springer}
}

@inproceedings{10.1145/3748699.3749798,
author = {Franco, Mirko and Grimm, Valentin and Herder, Eelco},
title = {Preventing Accidental Sharing of Misinformation Using Large Language Models},
year = {2025},
isbn = {9798400720895},
publisher = {Association for Computing Machinery},
address = {New York, NY, USA},
url = {https://doi.org/10.1145/3748699.3749798},
doi = {10.1145/3748699.3749798},
booktitle = {Proceedings of the 2025 International Conference on Information Technology for Social Good},
pages = {244–252},
numpages = {9},
location = {
},
series = {GoodIT '25}
}

\appendix

\section{Ethics Statement}

In our research, we collected data including posts and user profiles from publicly available sources (X, formerly Twitter) under the platform’s terms of service. All data utilised in our study have been strictly limited to academic research purposes. The data has not been used to target or identify any individual personally. We have taken rigorous measures to ensure the anonymity and privacy of the data subjects. We have not disclosed any of the data to third parties.

\section{Computational Resources}

Experiments in this study were conducted on a Linux-based workstation, equipped with an AMD Ryzen 9 9950X CPU, 64 GB of system memory, and an NVIDIA GeForce RTX 4090 GPU with 24 GB of memory.
Due to the large number of experimental runs and the computational cost, we additionally used a GPU server for the Qwen out-of-distribution experiments. This server is equipped with two NVIDIA A40 GPUs with 48 GB of memory per GPU, dual AMD EPYC 7443 processors with 48 cores and 96 threads in total, and 512 GB of system memory.
Our models are compatible with standard GPU-equipped servers. No conclusions in this study depend on specialised hardware.

\section{Hyper-parameters of the Decision Tree Model (Table~\ref{app:table_DT_parameters})}
\label{app:DT_parameters}

We searched for the optimal hyper-parameters of the Decision Tree (DT) model for each experiment. 
The hyper-parameters were determined on the validation set. 
Searched values are provided in Table~\ref{app:table_DT_parameters}, where values marked with * are most frequently used. The parameter \textit{scale\_pos\_weight} was only considered for imbalanced datasets (1:5 and 1:10).

\begin{table}[!t]
\centering
\renewcommand{\arraystretch}{1.2}
\caption{Hyper-parameters of the Decision Tree model}
\begin{tabular}{ll}
\toprule
Hyper-parameter & Searched values \\
\midrule
\textit{max\_depth} & 6, 7, 8*, 9, 10 \\
\textit{learning\_rate} & 0.3*, 0.35, 0.4 \\
\textit{n\_estimators} & 100*, 150, 200 \\
\textit{min\_child\_weight} & 1*, 2, 3 \\
\textit{subsample} & 0.8, 0.9, 1.0* \\
\textit{scale\_pos\_weight} & [1*, 2, 3, 4, 5] (1:5 dataset) \\
& [1.00*, 3.25, 5.50, 7.75, 10.00] \\
& (1:10 dataset) \\
\bottomrule
\end{tabular}
\label{app:table_DT_parameters}
\end{table}

\section{Serialisations of User-Related Features}
\label{app:example_serialisations}

DT and MLP models use engineered feature vectors directly, whereas BERT and Qwen require textual input. To enable BERT and Qwen to use structured user-related features, we follow the serialisation strategy in \citet{hegselmann2023tabllm}, which converts tabular feature values into natural-language descriptions.
Specifically, for each reposting instance, the sender and recipient values are presented together in the same sentence, so that the language model can process their relationship within a single textual input.
For example, the serialised user-related features for one reposting instance are represented as follows:

\texttt{Account age is 4042 and 55. Follower count is 3774 and 915. Following count is 4999 and 1347. Post count is 20162 and 541. Listed count is 12 and 0. Posting activity is 0.01 and 0. Follower growth is 0.93 and 16.51. Following growth is 1.24 and 24.31. Post growth is 4.99 and 9.76. Listed growth is 0 and 0. Verification is 0 and 0. URL is 1 and 0. LeaderRank is 7.5e-11 and 2.78e-10. In-degree is 0 and 0. Following relation is 1 and 1. Historical post count is 50 and 50. Original post percentage is 2\% and 2\%. Retweet percentage is 70\% and 44\%. Quote percentage is 0\% and 2\%. Reply percentage is 28\% and 52\%. Repost percentage is 98\% and 98\%. Post interval is 0.11 and 0.1. Retweet rate is 1.87 and 0.32. Quote rate is 0.07 and 0.04. Reply rate is 0.53 and 0.04. Like rate is 3.4 and 1.5. Dominant topic is news \& social concern and news \& social concern, strength is 0.9 and 0.63. Topic count is 1.26 and 1.08. Character count is 116.67 and 60.4. Word count is 21.79 and 11.4. Grammar score is 0.91 and 0.87 for words, 0.4 and 0.45 for sentences. Polarity is -0.08 and 0.05. Subjectivity is 0.28 and 0.29. Irony is 0.42 and 0.36. Offensiveness is 0.21 and 0.17. Gender is 0.81 and 0.64. Readability is 64.35 and 70.21. Sentiment is negative and positive, strength is -0.14 and 0.05. Emotion is others and others, strength is 0.54 and 0.54. Hate speech intensity is 0.04 and 0.04. Mention count is 0 and 0, percentage is 0\% and 0\%. Repost latency is 0. Topical similarity is 0.1.}

In this example, paired values correspond to the sender and recipient, respectively, unless the feature is defined for the sender--recipient pair as a whole, such as repost latency or topical similarity.

\section{Qwen Prompt}
\label{app:qwen_prompt}

In our paper, Qwen is used as a generative language-model-based classifier, which needs a system prompt to specify the task format and constrain the output space.
The prompt defines reposting prediction as a binary classification task, specifies the meaning of the two labels, and instructs the model to return only one label.
This is necessary because decoder-only language models generate text autoregressively; without an explicit instruction, the model may produce explanations or other text in addition to the predicted label.

The system prompt used in our experiments is:

\texttt{You are a binary classifier for repost prediction on X.
Predict whether the recipient will repost the post from the sender.
Label definitions: 0 = not reposted, 1 = reposted.
Return exactly one label: 0 or 1.
Do not explain your answer.}

BERT is an encoder-based classifier, and the output format is determined by the classification head, which does not need a system prompt.

\section{Sensitivity to Qwen Model Size (Table~\ref{app:tab_qwen_varients})}
\label{app:qwen_varients}

In the main experiments, we use \texttt{Qwen3-0.6B} as a lightweight and reproducible representative of the Qwen3 model family.
To examine whether the choice of model size affects the results, we additionally evaluate larger Qwen variants, including \texttt{Qwen3-4B} and \texttt{Qwen3-1.7B}, in the mixed-hashtag random-split experiment.
As shown in Table~\ref{app:tab_qwen_varients}, the results are close across different Qwen model sizes.
Therefore, we use \texttt{Qwen3-0.6B} in the main experiments as it is sufficient for our evaluation.

\begin{table}[!t]
\centering
\renewcommand{\arraystretch}{1.2}
\caption{In-Distribution Prediction Results for Mixed Hashtags Using Qwen Variants}
\begin{tabular}{l|ccc}
\hline
 & \multicolumn{3}{c}{\textbf{Setting}} \\ 
\textbf{Model} & \textbf{ALL} & \textbf{U} & \textbf{M} \\ 
\hline
\texttt{Qwen3-4B} & 0.869 & 0.856 & 0.747 \\
\texttt{Qwen3-1.7B} & 0.872 & 0.852 & 0.750 \\
\texttt{Qwen3-0.6B} & 0.873 & 0.850 & 0.748 \\
\hline
\end{tabular}
\label{app:tab_qwen_varients}
\end{table}

\section{Supplementary Results on 1:1, 1:5, and 1:10 Datasets (Table~\ref{app:table_results_1_1_1_10})}
\label{app:results_1_1_1_10}

In the main experiments, we report results on the 1:5 dataset, which is commonly used in reposting prediction experiments and serves as our primary dataset.
To examine whether the main findings are robust to different positive-to-negative ratios, we additionally evaluate the mixed-hashtag random-split prediction on two supplementary datasets with 1:1 and 1:10 positive-to-negative ratios. The 1:1 dataset contains one negative instance for each positive instance, while the 1:10 dataset contains ten negative instances for each positive instance.
The results are shown in Table \ref{app:table_results_1_1_1_10}. The conclusions drawn from the results are consistent with those presented in the main paper.

\begin{table}[!t]
\centering
\renewcommand{\arraystretch}{1.2}
\caption{In-Distribution Prediction Results for Mixed Hashtags Using the Decision Tree Model on 1:1, 1:5, and 1:10 Datasets. }
\begin{tabular}{cc|ccc}
\hline
& & \multicolumn{3}{c}{\textbf{Feature Setting}} \\ 
\textbf{Dataset} & \textbf{Model} & \textbf{ALL} & \textbf{U} & \textbf{M} \\ 
\hline
1:1 & DT
& \textbf{0.935}\scriptsize{$\pm$0.002} 
& 0.919\scriptsize{$\pm$0.003} 
& 0.871\scriptsize{$\pm$0.002} \\
1:5 & DT
& \textbf{0.884}\scriptsize{$\pm$0.002} 
& 0.852\scriptsize{$\pm$0.005} 
& 0.758\scriptsize{$\pm$0.002} \\
1:10 & DT
& \textbf{0.835}\scriptsize{$\pm$0.005} 
& 0.808\scriptsize{$\pm$0.004} 
& 0.666\scriptsize{$\pm$0.002} \\ 
\hline
\end{tabular}
\label{app:table_results_1_1_1_10}
\end{table}

\section{Supplementary Results on Two Negative Sampling Methods:  Most-Similar vs Random (Table~\ref{app:table_general})}

\begin{table}[!t]
\centering
\setlength{\tabcolsep}{1.7mm}
\renewcommand{\arraystretch}{1.2}
\caption{Supplementary Results under Alternative Negative Sampling}
\begin{tabular}{c|ccc|ccc|ccc}
\hline
Experiment: & \multicolumn{3}{c|}{Mixed-Hashtag} &\multicolumn{3}{c|}{Per-Hashtag} &\multicolumn{3}{c}{Out-of-distribution} \\ 
Model\,-\,Feature Setting: & DT-\textbf{ALL} & DT-\textbf{U} & DT-\textbf{M} & DT-\textbf{ALL} & DT-\textbf{U} & DT-\textbf{M} & DT-\textbf{ALL} & DT-\textbf{U} & DT-\textbf{M} \\ 
\hline
Negative Sampling Method  & \multicolumn{9}{c}{F1 score} \\
\hline
Most Similar & \textbf{0.884} & 0.852 & 0.758 & \textbf{0.868} & 0.841 & 0.742 & 0.697 & \textbf{0.705} & 0.117 \\
Random & \textbf{0.910} & 0.896 & 0.770 & \textbf{0.896} & 0.889 & 0.745 & \textbf{0.782} & 0.781 & 0.105 \\
\hline
\multicolumn{10}{p{0.95\linewidth}}{\footnotesize Note: This table shows the prediction results using the Decision Tree model with the 3 feature settings on two 1:5 datasets sampled with different negative sampling methods: (1) Most Similar, and (2) Random. Results in the main text are based on the 1:5 datasets with the most-similar negative sampling.}
\end{tabular}
\label{app:table_general}
\end{table}

In addition to the main experiments reported in the paper, where negative instances are constructed by selecting the five most similar non-reposted cases for each positive instance, we also conduct supplementary experiments on a randomly-sampled 1:5 dataset, which serve as a robustness check and evaluate our models under a more general sampling strategy.
For convenience, we first define the set of posts in our raw data as $\mathcal{P}=\{(M_h, U_S)\}$ and the set of reposts $\mathcal{RP}=\{(M_h, U_S, U_R)\}$~\footnote{~Note: Reposts whose original posts are absent from the raw data are excluded.}. We further denote $\mathcal{U}$ as the set of users, which includes all $U_S$ from $\mathcal{P}$ and $U_R$ from $\mathcal{RP}$.
In the supplementary experiments, a negative instance $(M^*_h, U^*_S, U^*_R)$ is generated by randomly pairing a post $(M^*_h, U^*_S) \in \mathcal{P}$ and a user $U^*_R \in \mathcal{U}$, on the condition that $(M^*_h, U^*_S, U^*_R) \notin \mathcal{RP}$.
Therefore, a negative instance is generated independently of a specific positive instance and without similarity constraints, which is more general than the strategy used in the main experiments.

We apply this supplementary sampling strategy to the mixed-hashtag, per-hashtag, and out-of-distribution predictions.
The temporal prediction is not included because the randomly generated negative instances are constructed independently of any positive instance.
Consequently, unlike the negatives used in the main experiments, they do not inherit a natural temporal-window assignment from a corresponding positive instance, making them incompatible with the rolling-window temporal prediction.
The prediction results are shown in Table \ref{app:table_general}. The results and conclusions are consistent with those presented in the main paper.

\section{Detailed Results for In-Distribution Per-Hashtag Prediction (Table~\ref{app:tab_per_hashtag_results})}
\label{app:per_hashtag_results}

We provide the detailed results for per-hashtag prediction in Table \ref{app:tab_per_hashtag_results}.
The results and conclusions are consistent with those presented in the main paper: models using ALL or U significantly outperform M overall and for most individual hashtags (except \texttt{\#Supercup}, \texttt{\#IOS16}, and \texttt{\#LoveIsland} for s).

\newpage

\begin{table}[!t]
\centering
\renewcommand{\arraystretch}{1.2}
\caption{Detailed Results for In-Distribution Per-Hashtag Prediction}
\begin{tabular}{r||ccc|ccc}
\hline\hline
\textbf{Models:} & \multicolumn{3}{c|}{{\textbf{DT}}} & \multicolumn{3}{c}{{\textbf{MLP}}} \\ 
\textbf{Settings:} & \textbf{ALL} & \textbf{U} & \textbf{M} & \textbf{ALL} & \textbf{U} & \textbf{M} \\  
\hline\hline
Hashtag & \multicolumn{6}{c}{{F1 Score ($\mu$ $\pm$ $\sigma$)}} \\ 
\hline
\texttt{\#NHS}& \textbf{0.937}\scriptsize{$\pm$0.003}& 0.929\scriptsize{$\pm$0.003}& 0.827\scriptsize{$\pm$0.004}& \textbf{0.917}\scriptsize{$\pm$0.003}& 0.907\scriptsize{$\pm$0.005}& 0.817\scriptsize{$\pm$0.006}\\  
\texttt{\#BLM}& \textbf{0.934}\scriptsize{$\pm$0.007}& 0.927\scriptsize{$\pm$0.007}& 0.823\scriptsize{$\pm$0.006}& \textbf{0.901}\scriptsize{$\pm$0.006}& 0.891\scriptsize{$\pm$0.007}& 0.813\scriptsize{$\pm$0.011}\\  
\texttt{\#Brexit}& \textbf{0.897}\scriptsize{$\pm$0.006}& 0.878\scriptsize{$\pm$0.005}& 0.821\scriptsize{$\pm$0.008}& \textbf{0.866}\scriptsize{$\pm$0.007}& 0.841\scriptsize{$\pm$0.007}& 0.812\scriptsize{$\pm$0.008}\\  
\texttt{\#Supercup}& \textbf{0.890}\scriptsize{$\pm$0.009}& 0.877\scriptsize{$\pm$0.005}& 0.814\scriptsize{$\pm$0.008}& \textbf{0.860}\scriptsize{$\pm$0.010}& 0.854\scriptsize{$\pm$0.010}& 0.810\scriptsize{$\pm$0.009}\\  
\texttt{\#Olivianewtonjohn}& \textbf{0.879}\scriptsize{$\pm$0.011}& 0.866\scriptsize{$\pm$0.012}& 0.773\scriptsize{$\pm$0.010}& \textbf{0.855}\scriptsize{$\pm$0.019}& 0.843\scriptsize{$\pm$0.019}& 0.771\scriptsize{$\pm$0.013}\\  
\texttt{\#Monkeypox}& \textbf{0.878}\scriptsize{$\pm$0.006}& 0.850\scriptsize{$\pm$0.008}& 0.750\scriptsize{$\pm$0.013}& \textbf{0.836}\scriptsize{$\pm$0.011}& 0.807\scriptsize{$\pm$0.011}& 0.726\scriptsize{$\pm$0.011}\\  
\texttt{\#EnergyPrices}& \textbf{0.871}\scriptsize{$\pm$0.009}& 0.856\scriptsize{$\pm$0.012}& 0.730\scriptsize{$\pm$0.011}& \textbf{0.824}\scriptsize{$\pm$0.018}& 0.798\scriptsize{$\pm$0.015}& 0.719\scriptsize{$\pm$0.010}\\  
\texttt{\#IOS16}& \textbf{0.864}\scriptsize{$\pm$0.017}& 0.841\scriptsize{$\pm$0.009}& 0.736\scriptsize{$\pm$0.025}& \textbf{0.803}\scriptsize{$\pm$0.020}& 0.786\scriptsize{$\pm$0.017}& 0.724\scriptsize{$\pm$0.027}\\  
\texttt{\#Avengers}& \textbf{0.861}\scriptsize{$\pm$0.009}& 0.858\scriptsize{$\pm$0.007}& 0.557\scriptsize{$\pm$0.017}& \textbf{0.832}\scriptsize{$\pm$0.008}& 0.825\scriptsize{$\pm$0.010}& 0.524\scriptsize{$\pm$0.013}\\  
\texttt{\#Climatechange}& \textbf{0.850}\scriptsize{$\pm$0.008}& 0.844\scriptsize{$\pm$0.009}& 0.712\scriptsize{$\pm$0.012}& \textbf{0.793}\scriptsize{$\pm$0.014}& 0.782\scriptsize{$\pm$0.011}& 0.699\scriptsize{$\pm$0.018}\\  
\texttt{\#Covid}& \textbf{0.844}\scriptsize{$\pm$0.012}& 0.839\scriptsize{$\pm$0.012}& 0.652\scriptsize{$\pm$0.016}& \textbf{0.814}\scriptsize{$\pm$0.009}& 0.799\scriptsize{$\pm$0.014}& 0.632\scriptsize{$\pm$0.025}\\  
\texttt{\#CostofLivingCrisis}& \textbf{0.835}\scriptsize{$\pm$0.011}& 0.820\scriptsize{$\pm$0.009}& 0.743\scriptsize{$\pm$0.013}& \textbf{0.812}\scriptsize{$\pm$0.013}& 0.793\scriptsize{$\pm$0.011}& 0.733\scriptsize{$\pm$0.016}\\  
\texttt{\#LoveIsland}& \textbf{0.808}\scriptsize{$\pm$0.007}& 0.668\scriptsize{$\pm$0.016}& 0.752\scriptsize{$\pm$0.006}& \textbf{0.764}\scriptsize{$\pm$0.007}& 0.656\scriptsize{$\pm$0.008}& 0.740\scriptsize{$\pm$0.009}\\  
\texttt{\#UkraineWar}& \textbf{0.797}\scriptsize{$\pm$0.008}& 0.717\scriptsize{$\pm$0.014}& 0.695\scriptsize{$\pm$0.020}& \textbf{0.737}\scriptsize{$\pm$0.015}& 0.689\scriptsize{$\pm$0.015}& 0.677\scriptsize{$\pm$0.024}\\  
\hline
Overall F1 Score ($\bar\mu \pm \bar\sigma$) & \textbf{0.868}\scriptsize{$\pm$0.040} & 0.841\scriptsize{$\pm$0.069} & 0.742\scriptsize{$\pm$0.073} & \textbf{0.829}\scriptsize{$\pm$0.049}& 0.805\scriptsize{$\pm$0.067}& 0.728\scriptsize{$\pm$0.080}\\  
\hline\hline
\textbf{Models:} & \multicolumn{3}{c|}{\textbf{BERT}} & \multicolumn{3}{c}{\textbf{Qwen}}\\ 
\textbf{Settings:} & \textbf{ALL} & \textbf{U} & \textbf{M} & \textbf{ALL} & \textbf{U} & \textbf{M} \\   
\hline\hline
Hashtag & \multicolumn{6}{c}{{F1 Score ($\mu$ $\pm$ $\sigma$)}} \\ 
\hline
\texttt{\#NHS}& 0.888\scriptsize{$\pm$0.005}& \textbf{0.902}\scriptsize{$\pm$0.016}& 0.818\scriptsize{$\pm$0.005}& \textbf{0.927}\scriptsize{$\pm$0.006}& 0.921\scriptsize{$\pm$0.005}& 0.817\scriptsize{$\pm$0.005}\\  
\texttt{\#BLM}& 0.871\scriptsize{$\pm$0.019}& \textbf{0.879}\scriptsize{$\pm$0.017}& 0.813\scriptsize{$\pm$0.008}& \textbf{0.903}\scriptsize{$\pm$0.003}& 0.898\scriptsize{$\pm$0.011}& 0.810\scriptsize{$\pm$0.006}\\  
\texttt{\#Brexit}& \textbf{0.826}\scriptsize{$\pm$0.017}& 0.811\scriptsize{$\pm$0.009}& 0.807\scriptsize{$\pm$0.017}& \textbf{0.865}\scriptsize{$\pm$0.019}& 0.859\scriptsize{$\pm$0.015}& 0.815\scriptsize{$\pm$0.009}\\  
\texttt{\#Supercup}& \textbf{0.810}\scriptsize{$\pm$0.010}& 0.805\scriptsize{$\pm$0.022}& 0.806\scriptsize{$\pm$0.010}& \textbf{0.838}\scriptsize{$\pm$0.005}& 0.815\scriptsize{$\pm$0.017}& 0.808\scriptsize{$\pm$0.006}\\  
\texttt{\#Olivianewtonjohn}& 0.778\scriptsize{$\pm$0.048}& \textbf{0.839}\scriptsize{$\pm$0.015}& 0.746\scriptsize{$\pm$0.024}& \textbf{0.862}\scriptsize{$\pm$0.013}& 0.843\scriptsize{$\pm$0.013}& 0.730\scriptsize{$\pm$0.034}\\  
\texttt{\#Monkeypox}& \textbf{0.816}\scriptsize{$\pm$0.027}& 0.808\scriptsize{$\pm$0.014}& 0.735\scriptsize{$\pm$0.014}& \textbf{0.844}\scriptsize{$\pm$0.006}& 0.815\scriptsize{$\pm$0.010}& 0.726\scriptsize{$\pm$0.014}\\  
\texttt{\#EnergyPrices}& \textbf{0.783}\scriptsize{$\pm$0.028}& 0.770\scriptsize{$\pm$0.038}& 0.707\scriptsize{$\pm$0.023}& \textbf{0.824}\scriptsize{$\pm$0.018}& 0.801\scriptsize{$\pm$0.020}& 0.707\scriptsize{$\pm$0.027}\\  
\texttt{\#IOS16}& 0.706\scriptsize{$\pm$0.038}& 0.720\scriptsize{$\pm$0.074}& \textbf{0.724}\scriptsize{$\pm$0.021}& \textbf{0.786}\scriptsize{$\pm$0.021}& 0.747\scriptsize{$\pm$0.013}& 0.702\scriptsize{$\pm$0.018}\\  
\texttt{\#Avengers}& \textbf{0.800}\scriptsize{$\pm$0.020}& 0.798\scriptsize{$\pm$0.025}& 0.530\scriptsize{$\pm$0.022}& \textbf{0.831}\scriptsize{$\pm$0.008}& 0.821\scriptsize{$\pm$0.007}& 0.514\scriptsize{$\pm$0.014}\\  
\texttt{\#Climatechange}& 0.761\scriptsize{$\pm$0.035}& \textbf{0.761}\scriptsize{$\pm$0.064}& 0.696\scriptsize{$\pm$0.018}& \textbf{0.800}\scriptsize{$\pm$0.010}& 0.769\scriptsize{$\pm$0.023}& 0.689\scriptsize{$\pm$0.019}\\
\texttt{\#Covid}& \textbf{0.788}\scriptsize{$\pm$0.020}& 0.771\scriptsize{$\pm$0.034}& 0.632\scriptsize{$\pm$0.018}& \textbf{0.798}\scriptsize{$\pm$0.017}& 0.764\scriptsize{$\pm$0.022}& 0.611\scriptsize{$\pm$0.032}\\  
\texttt{\#CostofLivingCrisis}& 0.742\scriptsize{$\pm$0.031}& \textbf{0.769}\scriptsize{$\pm$0.016}& 0.722\scriptsize{$\pm$0.027}& \textbf{0.814}\scriptsize{$\pm$0.011}& 0.784\scriptsize{$\pm$0.013}& 0.718\scriptsize{$\pm$0.020}\\  
\texttt{\#LoveIsland}& 0.739\scriptsize{$\pm$0.017}& 0.583\scriptsize{$\pm$0.096}& \textbf{0.739}\scriptsize{$\pm$0.013}& \textbf{0.748}\scriptsize{$\pm$0.007}& 0.639\scriptsize{$\pm$0.006}& 0.723\scriptsize{$\pm$0.013}\\  
\texttt{\#UkraineWar}& \textbf{0.694}\scriptsize{$\pm$0.020}& 0.652\scriptsize{$\pm$0.025}& 0.679\scriptsize{$\pm$0.017}& \textbf{0.718}\scriptsize{$\pm$0.011}& 0.656\scriptsize{$\pm$0.010}& 0.651\scriptsize{$\pm$0.024}\\  
\hline
Overall F1 Score ($\bar\mu \pm \bar\sigma$) & 0.786\scriptsize{$\pm$0.060} & 0.776\scriptsize{$\pm$0.090} & 0.725\scriptsize{$\pm$0.078} & \textbf{0.826}\scriptsize{$\pm$0.055}& 0.795\scriptsize{$\pm$0.078}& 0.716\scriptsize{$\pm$0.085}\\  
\hline\hline
\end{tabular}
\label{app:tab_per_hashtag_results}
\end{table}

\section{Detailed Results for In-Distribution Temporal Prediction (Table~\ref{app:tab_temporal_results})}
\label{app:temporal_results}

We provide the detailed results for temporal prediction in Table \ref{app:tab_temporal_results}. The results and conclusions are consistent with those in the main text. 
Both paired t-tests and Wilcoxon signed-rank tests confirm (with p-values $<$ 0.05) that models using ALL or U significantly outperform M overall and for most hashtags (except \texttt{\#Supercup} for DT, \texttt{\#Olivianewtonjohn}, \texttt{\#Supercup}, \texttt{\#IOS16}, \texttt{\#CostofLivingCrisis}, and \texttt{\#UkraineWar} for BERT, and \texttt{\#LoveIsland} for Qwen).

\newpage

\begin{table}[!t]
\centering
\renewcommand{\arraystretch}{1.2}
\caption{Detailed Results for In-Distribution Temporal Prediction}
\begin{tabular}{r||ccc|ccc}
\hline\hline
\textbf{Models:} & \multicolumn{3}{c|}{{\textbf{DT}}} & \multicolumn{3}{c}{{\textbf{MLP}}} \\ 
\textbf{Settings:} & \textbf{ALL} & \textbf{U} & \textbf{M} & \textbf{ALL} & \textbf{U} & \textbf{M} \\  
\hline\hline
Hashtag & \multicolumn{6}{c}{{F1 Score ($\mu$ $\pm$ $\sigma$)}} \\ 
\hline
\texttt{\#NHS}& \textbf{0.896}\scriptsize{$\pm$0.084}& 0.889\scriptsize{$\pm$0.086}& 0.679\scriptsize{$\pm$0.261}& 0.860\scriptsize{$\pm$0.112}& \textbf{0.870}\scriptsize{$\pm$0.102}& 0.710\scriptsize{$\pm$0.260}\\  
\texttt{\#Olivianewtonjohn}& \textbf{0.864}\scriptsize{$\pm$0.075}& 0.854\scriptsize{$\pm$0.070}& 0.690\scriptsize{$\pm$0.140}& 0.808\scriptsize{$\pm$0.093}& \textbf{0.834}\scriptsize{$\pm$0.082}& 0.641\scriptsize{$\pm$0.148}\\  
\texttt{\#BLM}& \textbf{0.849}\scriptsize{$\pm$0.067}& 0.831\scriptsize{$\pm$0.058}& 0.488\scriptsize{$\pm$0.228}& 0.761\scriptsize{$\pm$0.095}& \textbf{0.774}\scriptsize{$\pm$0.073}& 0.483\scriptsize{$\pm$0.260}\\ 
\texttt{\#Monkeypox}& \textbf{0.834}\scriptsize{$\pm$0.056}& 0.815\scriptsize{$\pm$0.057}& 0.597\scriptsize{$\pm$0.096}& \textbf{0.768}\scriptsize{$\pm$0.050}& 0.764\scriptsize{$\pm$0.067}& 0.504\scriptsize{$\pm$0.128}\\
\texttt{\#Covid}& 0.829\scriptsize{$\pm$0.049}& \textbf{0.831}\scriptsize{$\pm$0.043}& 0.451\scriptsize{$\pm$0.065}& \textbf{0.772}\scriptsize{$\pm$0.040}& 0.765\scriptsize{$\pm$0.048}& 0.443\scriptsize{$\pm$0.071}\\
\texttt{\#IOS16}& \textbf{0.824}\scriptsize{$\pm$0.076}& 0.788\scriptsize{$\pm$0.080}& 0.514\scriptsize{$\pm$0.318}& 0.741\scriptsize{$\pm$0.119}& \textbf{0.760}\scriptsize{$\pm$0.085}& 0.484\scriptsize{$\pm$0.223}\\ 
\texttt{\#Brexit}& \textbf{0.821}\scriptsize{$\pm$0.063}& 0.812\scriptsize{$\pm$0.069}& 0.625\scriptsize{$\pm$0.160}& 0.764\scriptsize{$\pm$0.062}& \textbf{0.765}\scriptsize{$\pm$0.067}& 0.613\scriptsize{$\pm$0.156}\\
\texttt{\#Climatechange}& \textbf{0.814}\scriptsize{$\pm$0.058}& 0.789\scriptsize{$\pm$0.046}& 0.560\scriptsize{$\pm$0.053}& \textbf{0.764}\scriptsize{$\pm$0.050}& 0.760\scriptsize{$\pm$0.055}& 0.533\scriptsize{$\pm$0.063}\\  
\texttt{\#CostofLivingCrisis}& \textbf{0.801}\scriptsize{$\pm$0.096}& 0.788\scriptsize{$\pm$0.083}& 0.632\scriptsize{$\pm$0.213}& \textbf{0.761}\scriptsize{$\pm$0.124}& 0.740\scriptsize{$\pm$0.110}& 0.622\scriptsize{$\pm$0.145}\\ 
\texttt{\#Supercup}& 0.799\scriptsize{$\pm$0.178}& \textbf{0.803}\scriptsize{$\pm$0.180}& 0.689\scriptsize{$\pm$0.240}& \textbf{0.785}\scriptsize{$\pm$0.172}& 0.752\scriptsize{$\pm$0.198}& 0.707\scriptsize{$\pm$0.233}\\   
\texttt{\#Avengers}& \textbf{0.793}\scriptsize{$\pm$0.085}& 0.791\scriptsize{$\pm$0.092}& 0.385\scriptsize{$\pm$0.185}& 0.717\scriptsize{$\pm$0.081}& \textbf{0.719}\scriptsize{$\pm$0.080}& 0.289\scriptsize{$\pm$0.187}\\  
\texttt{\#UkraineWar}& \textbf{0.779}\scriptsize{$\pm$0.070}& 0.670\scriptsize{$\pm$0.104}& 0.628\scriptsize{$\pm$0.123}& \textbf{0.721}\scriptsize{$\pm$0.078}& 0.717\scriptsize{$\pm$0.055}& 0.582\scriptsize{$\pm$0.133}\\  
\texttt{\#EnergyPrices}& \textbf{0.760}\scriptsize{$\pm$0.062}& 0.746\scriptsize{$\pm$0.058}& 0.490\scriptsize{$\pm$0.230}& \textbf{0.690}\scriptsize{$\pm$0.085}& 0.686\scriptsize{$\pm$0.064}& 0.461\scriptsize{$\pm$0.204}\\  
\texttt{\#LoveIsland}& \textbf{0.749}\scriptsize{$\pm$0.050}& 0.622\scriptsize{$\pm$0.049}& 0.686\scriptsize{$\pm$0.110}& \textbf{0.730}\scriptsize{$\pm$0.040}& 0.638\scriptsize{$\pm$0.074}& 0.681\scriptsize{$\pm$0.074}\\ 
\hline
Overall F1 Score ($\bar\mu \pm \bar\sigma$) & \textbf{0.815}\scriptsize{$\pm$0.091} & 0.788\scriptsize{$\pm$0.107} & 0.579\scriptsize{$\pm$0.212} & \textbf{0.760}\scriptsize{$\pm$0.101} & 0.753\scriptsize{$\pm$0.105} & 0.554\scriptsize{$\pm$0.210} \\  
\hline\hline
\textbf{Models:} & \multicolumn{3}{c|}{\textbf{BERT}} & \multicolumn{3}{c}{\textbf{Qwen}}\\ 
\textbf{Settings:} & \textbf{ALL} & \textbf{U} & \textbf{M} & \textbf{ALL} & \textbf{U} & \textbf{M} \\  
\hline\hline
Hashtag & \multicolumn{6}{c}{{F1 Score ($\mu$ $\pm$ $\sigma$)}} \\ 
\hline
\texttt{\#NHS}& 0.748\scriptsize{$\pm$0.220}& \textbf{0.830}\scriptsize{$\pm$0.129}& 0.639\scriptsize{$\pm$0.297}& 0.793\scriptsize{$\pm$0.186}& \textbf{0.847}\scriptsize{$\pm$0.124}& 0.645\scriptsize{$\pm$0.315}\\  
\texttt{\#Olivianewtonjohn}& 0.677\scriptsize{$\pm$0.163}& \textbf{0.721}\scriptsize{$\pm$0.057}& 0.641\scriptsize{$\pm$0.141}& 0.802\scriptsize{$\pm$0.098}& \textbf{0.802}\scriptsize{$\pm$0.097}& 0.562\scriptsize{$\pm$0.176}\\  
\texttt{\#BLM}& 0.543\scriptsize{$\pm$0.196}& \textbf{0.746}\scriptsize{$\pm$0.076}& 0.509\scriptsize{$\pm$0.240}& 0.745\scriptsize{$\pm$0.152}& \textbf{0.798}\scriptsize{$\pm$0.085}& 0.410\scriptsize{$\pm$0.285}\\ 
\texttt{\#Monkeypox}& 0.533\scriptsize{$\pm$0.061}& \textbf{0.730}\scriptsize{$\pm$0.064}& 0.494\scriptsize{$\pm$0.133}& 0.749\scriptsize{$\pm$0.069}& \textbf{0.758}\scriptsize{$\pm$0.050}& 0.444\scriptsize{$\pm$0.143}\\ 
\texttt{\#Covid}& 0.522\scriptsize{$\pm$0.111}& \textbf{0.700}\scriptsize{$\pm$0.050}& 0.324\scriptsize{$\pm$0.135}& 0.764\scriptsize{$\pm$0.051}& \textbf{0.765}\scriptsize{$\pm$0.046}& 0.396\scriptsize{$\pm$0.104}\\ 
\texttt{\#IOS16}& 0.586\scriptsize{$\pm$0.200}& \textbf{0.623}\scriptsize{$\pm$0.251}& 0.511\scriptsize{$\pm$0.316}& \textbf{0.734}\scriptsize{$\pm$0.116}& 0.714\scriptsize{$\pm$0.068}& 0.510\scriptsize{$\pm$0.276}\\ 
\texttt{\#Brexit}& 0.631\scriptsize{$\pm$0.140}& \textbf{0.722}\scriptsize{$\pm$0.092}& 0.606\scriptsize{$\pm$0.160}& 0.725\scriptsize{$\pm$0.085}& \textbf{0.749}\scriptsize{$\pm$0.094}& 0.604\scriptsize{$\pm$0.154}\\      
\texttt{\#Climatechange}& 0.547\scriptsize{$\pm$0.043}& \textbf{0.662}\scriptsize{$\pm$0.129}& 0.544\scriptsize{$\pm$0.063}& 0.755\scriptsize{$\pm$0.067}& \textbf{0.756}\scriptsize{$\pm$0.058}& 0.533\scriptsize{$\pm$0.062}\\  
\texttt{\#CostofLivingCrisis}& 0.598\scriptsize{$\pm$0.155}& \textbf{0.624}\scriptsize{$\pm$0.228}& 0.566\scriptsize{$\pm$0.276}& 0.731\scriptsize{$\pm$0.130}& \textbf{0.753}\scriptsize{$\pm$0.081}& 0.572\scriptsize{$\pm$0.230}\\
\texttt{\#Supercup}& 0.723\scriptsize{$\pm$0.232}& \textbf{0.759}\scriptsize{$\pm$0.182}& 0.656\scriptsize{$\pm$0.278}& 0.754\scriptsize{$\pm$0.224}& \textbf{0.809}\scriptsize{$\pm$0.159}& 0.638\scriptsize{$\pm$0.291}\\
\texttt{\#Avengers}& 0.445\scriptsize{$\pm$0.279}& \textbf{0.591}\scriptsize{$\pm$0.186}& 0.238\scriptsize{$\pm$0.238}& 0.700\scriptsize{$\pm$0.079}& \textbf{0.730}\scriptsize{$\pm$0.085}& 0.264\scriptsize{$\pm$0.226}\\  
\texttt{\#UkraineWar}& 0.563\scriptsize{$\pm$0.220}& \textbf{0.614}\scriptsize{$\pm$0.059}& 0.590\scriptsize{$\pm$0.154}& \textbf{0.698}\scriptsize{$\pm$0.082}& 0.626\scriptsize{$\pm$0.093}& 0.603\scriptsize{$\pm$0.147}\\  
\texttt{\#EnergyPrices}& 0.472\scriptsize{$\pm$0.246}& \textbf{0.571}\scriptsize{$\pm$0.170}& 0.414\scriptsize{$\pm$0.264}& 0.606\scriptsize{$\pm$0.146}& \textbf{0.666}\scriptsize{$\pm$0.089}& 0.393\scriptsize{$\pm$0.282}\\  
\texttt{\#LoveIsland}& \textbf{0.699}\scriptsize{$\pm$0.058}& 0.576\scriptsize{$\pm$0.077}& 0.691\scriptsize{$\pm$0.087}& \textbf{0.707}\scriptsize{$\pm$0.081}& 0.581\scriptsize{$\pm$0.050}& 0.695\scriptsize{$\pm$0.069}\\  
\hline
Overall F1 Score ($\bar\mu \pm \bar\sigma$) & 0.592\scriptsize{$\pm$0.202}& \textbf{0.676}\scriptsize{$\pm$0.160}& 0.530\scriptsize{$\pm$0.248}& 0.733\scriptsize{$\pm$0.130}& \textbf{0.740}\scriptsize{$\pm$0.114}& 0.519\scriptsize{$\pm$0.244}  \\ 
\hline\hline
\end{tabular}
\label{app:tab_temporal_results}
\end{table}

\section{Temporal Prediction: Sliding and Accumulated Time Windows (Table~\ref{app:tab_accum_time_windows})}
\label{app:accum_time_windows}

In the main temporal prediction experiment, we use a sliding-window split, where a fixed number of recent time windows is used for training, and the next window is used for testing.
Specifically, three consecutive windows are used for training, and the following window is used for testing, sliding forward by one window at each step.
Here, we additionally report results under an accumulated-window split.
In this split, all available earlier windows are used for training, and the next window is used for testing. For example, windows 1--5 are used to predict window 6, rather than using only windows 3--5 as in the sliding-window split.
This comparison examines whether using a longer training history changes the relative performance of ALL, U, and M in temporal prediction.

As shown in Table \ref{app:tab_accum_time_windows}, the results of the two split strategies are very close, indicating that the main temporal prediction results are not sensitive to whether the training history is restricted to recent windows or accumulated from all earlier windows. The main patterns persist when the training history is accumulated over time.

\begin{table}[!t]
\centering
\renewcommand{\arraystretch}{1.2}
\caption{Temporal Prediction Performance Under Sliding and Accumulated Time Windows}
\begin{tabular}{c|ccc}
\hline
& \multicolumn{3}{c}{\textbf{Settings}} \\
\textbf{Temporal Split Strategy} & DT-\textbf{ALL} & DT-\textbf{U} & DT-\textbf{M} \\
\hline
Sliding & \textbf{0.815}\scriptsize{$\pm$0.091} & 0.788\scriptsize{$\pm$0.107} & 0.579\scriptsize{$\pm$0.212} \\
Accumulated & \textbf{0.819}\scriptsize{$\pm$0.089} & 0.783\scriptsize{$\pm$0.118} & 0.572\scriptsize{$\pm$0.203} \\
\hline
\end{tabular}
\label{app:tab_accum_time_windows}
\end{table}

\section{Random Guessing Prediction (Table~\ref{app:tab_random_guess})}
\label{app:sec_random_guess}

Table~\ref{app:tab_random_guess} shows the expected performance of different random guessing prediction strategies on the 1:5 dataset. 
\begin{itemize}
    \item \emph{Balanced prediction} randomly predicts positive and negative instances with equal probability, resulting in an F1 score of 0.250. 
    \item \emph{Matches true ratio} predicts positive and negative instances following the true 1:5 class distribution, producing an F1 score of 0.167. 
    \item \emph{All-positive prediction} predicts every instance as positive, achieving perfect recall but low precision. The F1 score is 0.286. 
    \item \emph{All-negative prediction} predicts every instance as negative. Since no instances are predicted as positive, precision is undefined, recall becomes 0.000, and the resulting F1 score is also 0.000, despite achieving the highest accuracy. 
\end{itemize}

\begin{table}[!t]
\centering
\caption{Expected F1 scores of Different Random Guessing Prediction Strategies on the 1:5 Dataset}
\renewcommand{\arraystretch}{1.2} 
\begin{tabular}{l|cccc}
\hline
\textbf{Prediction Strategy} & \textbf{Accuracy} & \textbf{Precision} & \textbf{Recall} & \textbf{F1 Score} \\
\hline
All-positive prediction & 0.167 & 0.167 & 1.000 & 0.286 \\
All-negative prediction & 0.833 & -- & 0.000 & 0.000 \\
Balanced prediction & 0.500 & 0.167 & 0.500 & 0.250 \\
Matches true ratio & 0.722 & 0.167 & 0.167 & 0.167 \\
\hline
\end{tabular}
\label{app:tab_random_guess}
\end{table}

\section{Additional Evaluation Metrics (Table~\ref{app:tab_gen_DT_other_metrics})}
\label{app:sec_gen_DT_other_metrics}

To further validate the conclusions drawn from the F1 scores in the main experiments, we additionally report 
precision and recall, alongside the F1 score, which are defined as follows:

\begin{equation}
\text{Precision} = \frac{TP}{TP + FP} \, ,
\end{equation}

\begin{equation}
\text{Recall} = \frac{TP}{TP + FN} \, ,
\end{equation}

\begin{equation}
\text{F1} = \frac{2 \times \text{Precision} \times \text{Recall}}
{\text{Precision} + \text{Recall}} \, ,
\end{equation}

where TP, TN, FP, and FN denote true positives, true negatives, false positives, and false negatives, respectively.

We focus on the out-of-distribution evaluation, as models using only post-related features exhibit a particularly severe performance collapse, making it important to examine whether this is driven by low precision, low recall, or both.

As shown in Table~\ref{app:tab_gen_DT_other_metrics}, the model using user-related features (DT-U) achieves the highest overall recall (as well as F1 score in Table~\ref{tab:table_4}), and the model using all features (DT-ALL) has the highest overall precision. DT-M, which uses only post-related features, performs substantially worse across all metrics. 

Overall, the additional evaluation metrics are consistent with the conclusions drawn from the F1 scores in the main experiments, further supporting the robustness of our findings that user-related features generalise substantially better than post-related features under unseen topics.

\begin{table}[!t]
\centering
\renewcommand{\arraystretch}{1.2}
\caption{Precision and Recall 
of the Decision Tree (DT) Model in Out-of-Distribution Prediction}
\begin{tabular}{r||ccc|ccc}
\hline\hline
\textbf{Metric:} & \multicolumn{3}{c|}{\textbf{Precision}} & \multicolumn{3}{c}{\textbf{Recall}} \\ 
\textbf{Settings:} & \textbf{DT-ALL} & \textbf{DT-U} & \textbf{DT-M} & \textbf{DT-ALL} & \textbf{DT-U} & \textbf{DT-M} \\  
\hline\hline
Hashtag & \multicolumn{6}{c}{{Metric Value ($\mu$ $\pm$ $\sigma$)}} \\ 
\hline
\texttt{\#BLM} 
& 0.849\scriptsize{$\pm$0.030}& \textbf{0.870}\scriptsize{$\pm$0.016}& 0.513\scriptsize{$\pm$0.294}& 0.779\scriptsize{$\pm$0.025}& \textbf{0.797}\scriptsize{$\pm$0.012}& 0.269\scriptsize{$\pm$0.189}\\  
\texttt{\#Covid} 
& \textbf{0.840}\scriptsize{$\pm$0.026}& 0.790\scriptsize{$\pm$0.023}& 0.224\scriptsize{$\pm$0.091}& 0.781\scriptsize{$\pm$0.015}& \textbf{0.787}\scriptsize{$\pm$0.018}& 0.062\scriptsize{$\pm$0.030}\\  
\texttt{\#NHS} 
& \textbf{0.757}\scriptsize{$\pm$0.070}& 0.737\scriptsize{$\pm$0.075}& 0.134\scriptsize{$\pm$0.112}& 0.746\scriptsize{$\pm$0.106}& \textbf{0.808}\scriptsize{$\pm$0.068}& 0.164\scriptsize{$\pm$0.186}\\  
\texttt{\#Monkeypox} 
& \textbf{0.775}\scriptsize{$\pm$0.024}& 0.666\scriptsize{$\pm$0.081}& 0.560\scriptsize{$\pm$0.105}& 0.699\scriptsize{$\pm$0.014}& \textbf{0.727}\scriptsize{$\pm$0.021}& 0.077\scriptsize{$\pm$0.020}\\  
\texttt{\#Climatechange} 
& \textbf{0.665}\scriptsize{$\pm$0.085}& 0.648\scriptsize{$\pm$0.040}& 0.248\scriptsize{$\pm$0.138}& 0.827\scriptsize{$\pm$0.026}& \textbf{0.854}\scriptsize{$\pm$0.018}& 0.097\scriptsize{$\pm$0.103}\\  
\texttt{\#Brexit} 
& \textbf{0.710}\scriptsize{$\pm$0.025}& 0.691\scriptsize{$\pm$0.025}& 0.283\scriptsize{$\pm$0.198}& 0.751\scriptsize{$\pm$0.040}& \textbf{0.809}\scriptsize{$\pm$0.023}& 0.030\scriptsize{$\pm$0.019}\\  
\texttt{\#IOS16} 
& 0.683\scriptsize{$\pm$0.038}& \textbf{0.695}\scriptsize{$\pm$0.025}& 0.215\scriptsize{$\pm$0.228}& 0.769\scriptsize{$\pm$0.024}& \textbf{0.813}\scriptsize{$\pm$0.022}& 0.054\scriptsize{$\pm$0.084}\\  
\texttt{\#Supercup} 
& 0.572\scriptsize{$\pm$0.029}& \textbf{0.631}\scriptsize{$\pm$0.038}& 0.245\scriptsize{$\pm$0.194}& \textbf{0.931}\scriptsize{$\pm$0.014}& 0.900\scriptsize{$\pm$0.050}& 0.202\scriptsize{$\pm$0.198}\\  
\texttt{\#Olivianewtonjohn} 
& 0.587\scriptsize{$\pm$0.080}& \textbf{0.727}\scriptsize{$\pm$0.062}& 0.178\scriptsize{$\pm$0.182}& \textbf{0.783}\scriptsize{$\pm$0.026}& 0.775\scriptsize{$\pm$0.021}& 0.065\scriptsize{$\pm$0.053}\\  
\texttt{\#CostofLivingCrisis} 
& 0.570\scriptsize{$\pm$0.048}& \textbf{0.579}\scriptsize{$\pm$0.044}& 0.114\scriptsize{$\pm$0.096}& 0.766\scriptsize{$\pm$0.020}& \textbf{0.789}\scriptsize{$\pm$0.023}& 0.030\scriptsize{$\pm$0.021}\\  
\texttt{\#Avengers} 
& 0.591\scriptsize{$\pm$0.052}& \textbf{0.595}\scriptsize{$\pm$0.039}& 0.153\scriptsize{$\pm$0.095}& 0.676\scriptsize{$\pm$0.055}& \textbf{0.756}\scriptsize{$\pm$0.018}& 0.057\scriptsize{$\pm$0.088}\\  
\texttt{\#UkraineWar} 
& \textbf{0.508}\scriptsize{$\pm$0.038}& 0.420\scriptsize{$\pm$0.045}& 0.159\scriptsize{$\pm$0.082}& 0.792\scriptsize{$\pm$0.025}& \textbf{0.857}\scriptsize{$\pm$0.029}& 0.048\scriptsize{$\pm$0.038}\\  
\texttt{\#EnergyPrices} 
& \textbf{0.821}\scriptsize{$\pm$0.020}& 0.798\scriptsize{$\pm$0.012}& 0.218\scriptsize{$\pm$0.031}& 0.474\scriptsize{$\pm$0.042}& \textbf{0.480}\scriptsize{$\pm$0.036}& 0.136\scriptsize{$\pm$0.041}\\  
\texttt{\#LoveIsland} 
& \textbf{0.532}\scriptsize{$\pm$0.023}& 0.473\scriptsize{$\pm$0.045}& 0.276\scriptsize{$\pm$0.144}& 0.675\scriptsize{$\pm$0.055}& \textbf{0.743}\scriptsize{$\pm$0.062}& 0.027\scriptsize{$\pm$0.033}\\  
\hline
Overall ($\bar\mu \pm \bar\sigma$) 
& \textbf{0.676}\scriptsize{$\pm$0.123} & 0.666\scriptsize{$\pm$0.127} & 0.251\scriptsize{$\pm$0.202} & 0.746\scriptsize{$\pm$0.106} & \textbf{0.778}\scriptsize{$\pm$0.100} & 0.094\scriptsize{$\pm$0.123} \\  
\hline\hline
\end{tabular}

\label{app:tab_gen_DT_other_metrics}
\end{table}

\section{Analysis with Inverted Predictions (Table~\ref{app:tab_inverted_pred})}
\label{app:sec_inverted_pred}

\begin{table}[!t]
\centering
\renewcommand{\arraystretch}{1.2}
\caption{Performance of Original and Inverted Predictions in the Out-of-Distribution Experiment}
\begin{tabular}{r||cc|cc|cc}
\hline\hline
\textbf{Metric:} & \multicolumn{2}{c|}{\textbf{Precision}} & \multicolumn{2}{c|}{\textbf{Recall}} & \multicolumn{2}{c}{\textbf{F1 score}}\\
\textbf{Prediction Type:} & \textbf{Original} & \textbf{Inverted} & \textbf{Original} & \textbf{Inverted} & \textbf{Original} & \textbf{Inverted} \\
\hline\hline
Hashtag &\multicolumn{6}{c}{{Metric Value of DT-M ($\mu$ $\pm$ $\sigma$)}}\\ 
\hline
 \texttt{\#BLM} &     \textbf{0.513}\scriptsize{$\pm$0.294}&0.131\scriptsize{$\pm$0.030}&0.269\scriptsize{$\pm$0.189}&\textbf{0.731}\scriptsize{$\pm$0.189}&\textbf{0.347}\scriptsize{$\pm$0.232}& 0.223\scriptsize{$\pm$0.052}\\
 \texttt{\#Covid} &     \textbf{0.224}\scriptsize{$\pm$0.091}&0.165\scriptsize{$\pm$0.008}&0.062\scriptsize{$\pm$0.030}&\textbf{0.938}\scriptsize{$\pm$0.030}&0.095\scriptsize{$\pm$0.045}& \textbf{0.281}\scriptsize{$\pm$0.013}\\
 \texttt{\#NHS} &     0.134\scriptsize{$\pm$0.112}&\textbf{0.168}\scriptsize{$\pm$0.031}&0.164\scriptsize{$\pm$0.186}&\textbf{0.836}\scriptsize{$\pm$0.186}&0.143\scriptsize{$\pm$0.141}& \textbf{0.280}\scriptsize{$\pm$0.054}\\
 \texttt{\#Monkeypox} &     \textbf{0.560}\scriptsize{$\pm$0.105}&0.163\scriptsize{$\pm$0.002}&0.077\scriptsize{$\pm$0.020}&\textbf{0.923}\scriptsize{$\pm$0.020}&0.133\scriptsize{$\pm$0.026}& \textbf{0.277}\scriptsize{$\pm$0.004}\\
 \texttt{\#Climatechange} &     \textbf{0.248}\scriptsize{$\pm$0.138}&0.159\scriptsize{$\pm$0.012}&0.097\scriptsize{$\pm$0.103}&\textbf{0.903}\scriptsize{$\pm$0.103}&0.127\scriptsize{$\pm$0.112}& \textbf{0.270}\scriptsize{$\pm$0.022}\\
 \texttt{\#Brexit} &     \textbf{0.283}\scriptsize{$\pm$0.198}&0.165\scriptsize{$\pm$0.003}&0.030\scriptsize{$\pm$0.019}&\textbf{0.970}\scriptsize{$\pm$0.019}&0.054\scriptsize{$\pm$0.033}& \textbf{0.282}\scriptsize{$\pm$0.005}\\
 \texttt{\#IOS16} &     \textbf{0.215}\scriptsize{$\pm$0.228}&0.164\scriptsize{$\pm$0.014}&0.054\scriptsize{$\pm$0.084}&\textbf{0.946}\scriptsize{$\pm$0.084}&0.085\scriptsize{$\pm$0.124}& \textbf{0.279}\scriptsize{$\pm$0.023}\\
 \texttt{\#Supercup} &     \textbf{0.245}\scriptsize{$\pm$0.194}&0.153\scriptsize{$\pm$0.029}&0.202\scriptsize{$\pm$0.198}&\textbf{0.798}\scriptsize{$\pm$0.198}&0.201\scriptsize{$\pm$0.177}& \textbf{0.256}\scriptsize{$\pm$0.051}\\
 \texttt{\#Olivianewtonjohn} &     0.178\scriptsize{$\pm$0.182}&\textbf{0.187}\scriptsize{$\pm$0.040}&0.065\scriptsize{$\pm$0.053}&\textbf{0.935}\scriptsize{$\pm$0.053}&0.084\scriptsize{$\pm$0.082}& \textbf{0.310}\scriptsize{$\pm$0.055}\\
 \texttt{\#CostofLivingCrisis} &     0.114\scriptsize{$\pm$0.096}&\textbf{0.183}\scriptsize{$\pm$0.026}&0.030\scriptsize{$\pm$0.021}&\textbf{0.970}\scriptsize{$\pm$0.021}&0.036\scriptsize{$\pm$0.022}& \textbf{0.307}\scriptsize{$\pm$0.034}\\
 \texttt{\#Avengers} &     0.153\scriptsize{$\pm$0.095}&\textbf{0.168}\scriptsize{$\pm$0.005}&0.057\scriptsize{$\pm$0.088}&\textbf{0.943}\scriptsize{$\pm$0.088}&0.063\scriptsize{$\pm$0.065}& \textbf{0.285}\scriptsize{$\pm$0.010}\\
 \texttt{\#UkraineWar} &     0.159\scriptsize{$\pm$0.082}&\textbf{0.171}\scriptsize{$\pm$0.010}&0.048\scriptsize{$\pm$0.038}&\textbf{0.952}\scriptsize{$\pm$0.038}&0.066\scriptsize{$\pm$0.039}& \textbf{0.290}\scriptsize{$\pm$0.014}\\
 \texttt{\#EnergyPrices} &     \textbf{0.218}\scriptsize{$\pm$0.031}&0.162\scriptsize{$\pm$0.004}&0.136\scriptsize{$\pm$0.041}&\textbf{0.864}\scriptsize{$\pm$0.041}&0.164\scriptsize{$\pm$0.033}& \textbf{0.272}\scriptsize{$\pm$0.007}\\
 \texttt{\#LoveIsland} &     \textbf{0.276}\scriptsize{$\pm$0.144}&0.165\scriptsize{$\pm$0.003}&0.027\scriptsize{$\pm$0.033}&\textbf{0.973}\scriptsize{$\pm$0.033}&0.045\scriptsize{$\pm$0.048}& \textbf{0.282}\scriptsize{$\pm$0.006}\\
\hline
 Overall ($\bar\mu \pm \bar\sigma$) & \textbf{0.251}\scriptsize{$\pm$0.202} & 0.165\scriptsize{$\pm$0.024} & 0.094\scriptsize{$\pm$0.123} & \textbf{0.906}\scriptsize{$\pm$0.123} & 0.117\scriptsize{$\pm$0.131} & \textbf{0.278}\scriptsize{$\pm$0.038} \\
\hline\hline
\end{tabular}

\label{app:tab_inverted_pred}
\end{table}

We additionally consider \emph{inverted prediction}, where all predicted labels are inverted, i.e., positive predictions become negative and negative predictions become positive.
Consequently, true positives become false negatives, false negatives become true positives, true negatives become false positives, and false positives become true negatives.
Under inverted prediction, the confusion matrix changes from

\[
\begin{bmatrix}
TN & FP \\
FN & TP
\end{bmatrix}
\]

to

\[
\begin{bmatrix}
FP & TN \\
TP & FN
\end{bmatrix} \, .
\]

The recall under original prediction is
\[
R = \frac{TP}{TP + FN} \, ,
\]

and the recall under inverted prediction becomes

\[
R' = \frac{FN}{FN + TP} = 1 - R \, .
\]

However, precision and F1 score do not have such simple symmetric relationships because they additionally depend on the number of true negatives.
For example, the precision under original prediction is
\[
P = \frac{TP}{TP + FP} \, ,
\]

The precision under inverted prediction becomes

\[
P' = \frac{FN}{FN + TN} \neq 1 - P \, .
\]

Inverted prediction can be used to check whether the poor performance of post-based models under out-of-distribution evaluation is simply because their predictions are systematically reversed.
If so, flipping all predicted labels should substantially improve their performance.

As shown in Table~\ref{app:tab_inverted_pred}, we have tested the inverted prediction on DT-M. 
Although inverted prediction greatly increases recall, the precision remains low. 
The overall F1 score only increases from 0.117 to 0.278, which still indicates limited predictive ability.
This suggests that the poor performance of DT-M under out-of-distribution evaluation is not simply due to learning the opposite label pattern, but reveals the limited predictive power of post-related features under unseen topics.

\begin{figure}[!t]
\centering
\includegraphics[width=\textwidth]{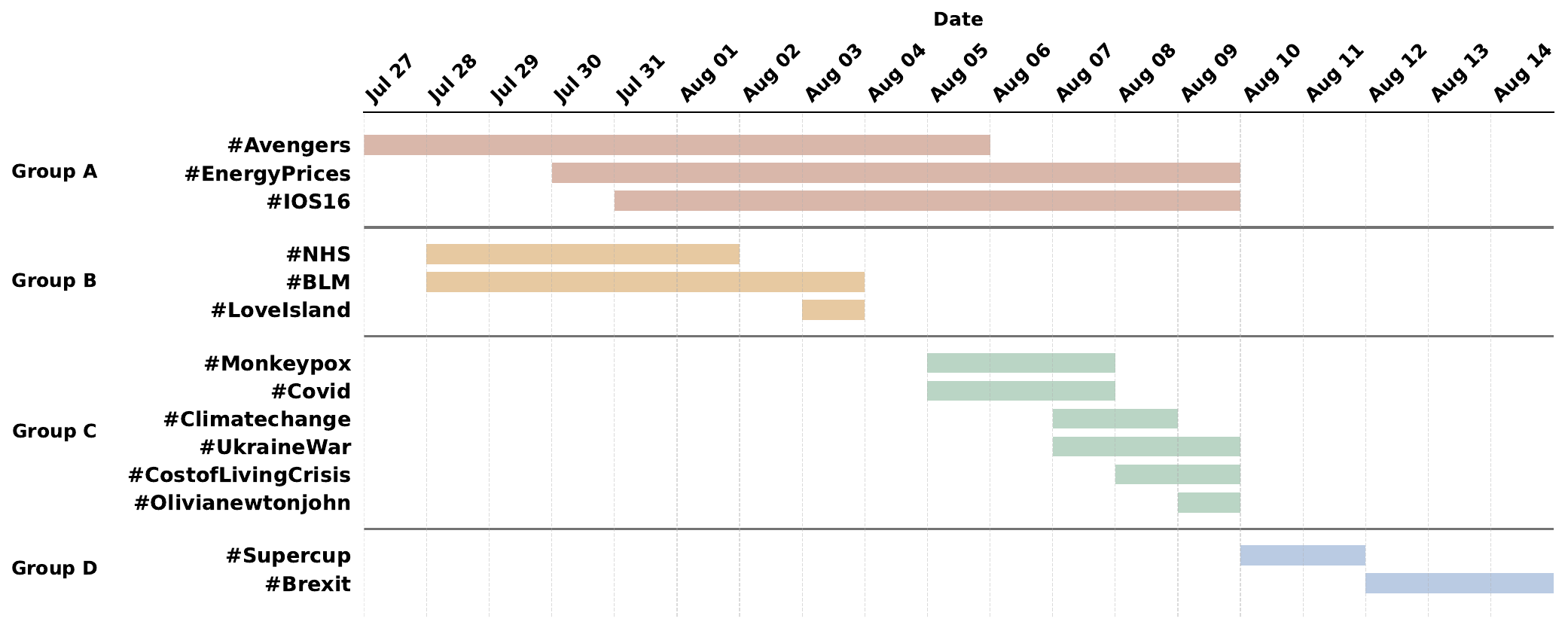}
\caption{Timelines of the studied hashtags.} 
\label{app:fig_hashtag_timelines} 
\end{figure}

\section{Temporally Constrained Out-of-Distribution Prediction (Figure~\ref{app:fig_hashtag_timelines} and Table~\ref{app:tab_hashtag_timelines})}

As an additional robustness check, we also conduct a temporally constrained out-of-distribution experiment, where earlier hashtags are used to predict later hashtags. The hashtag timelines are shown in Figure~\ref{app:fig_hashtag_timelines}, and results are presented in Table~\ref{app:tab_hashtag_timelines}.
The results and conclusions are consistent with those presented in the main paper.

\begin{table}[!t]
\centering
\renewcommand{\arraystretch}{1.2}
\caption{Out-of-distribution prediction under temporal constraints, using earlier hashtags to predict later hashtags.}
\begin{tabular}{cc|ccc}
\hline
& & \multicolumn{3}{c}{\textbf{Settings}} \\
& & DT-\textbf{ALL} & DT-\textbf{U} & DT-\textbf{M} \\
\hline
Training hashtags & Test hashtags & \multicolumn{3}{c}{{F1 Score ($\mu$ $\pm$ $\sigma$)}}\\
\hline
Group A+B+C & Group D & 0.720\scriptsize{$\pm$0.009} & \textbf{0.733}\scriptsize{$\pm$0.012} & 0.052\scriptsize{$\pm$0.027} \\
\hline
Group B & Group C & 0.626\scriptsize{$\pm$0.012} & \textbf{0.650}\scriptsize{$\pm$0.008} & 0.108\scriptsize{$\pm$0.015}\\
Group C & Group D & 0.706\scriptsize{$\pm$0.017} & \textbf{0.731}\scriptsize{$\pm$0.008} & 0.190\scriptsize{$\pm$0.073} \\
Group B & Group D & \textbf{0.734}\scriptsize{$\pm$0.008} & 0.724\scriptsize{$\pm$0.012} & 0.076\scriptsize{$\pm$0.072} \\
\hline
\multicolumn{5}{p{0.6\linewidth}}{\footnotesize
\textbf{Group A}: \#Avengers, \#EnergyPrices, \#IOS16;
\textbf{Group B}: \#NHS, \#BLM, \#LoveIsland;
\textbf{Group C}: \#Monkeypox, \#Covid, \#Climatechange, \#UkraineWar, \#CostofLivingCrisis, \#Olivianewtonjohn;
\textbf{Group D}: \#Supercup, \#Brexit.
}
\end{tabular}
\label{app:tab_hashtag_timelines}
\end{table}

\section{Full Lists of Features (Table~\ref{app:Mfeatures}, \ref{app:Ufeatures}, and \ref{app:HUfeatures})}
\label{app:features}

The following tables show the full lists of Message (M) features, User Profile (U-P) features, and User Historical Action (U-HA) features. User Historical Message (U-HM) features are the aggregated Message features (M) of a sender and those of a recipient, as well as an additional feature measuring topic similarity between the two users.
 
The following language models and tools are used for calculating M features.    
\begin{itemize}
    \item Topic features: Twitter-roBERTa-base\,\footnote{\,\url{https://huggingface.co/cardiffnlp}}, and Latent Dirichlet Allocation model from scikit-learn.
    \item Language features: language-tool-python\,\footnote{\,\url{https://github.com/jxmorris12/language\_tool\_python}}, TextBlob\,\footnote{\,\url{https://github.com/sloria/textblob}}, and a Logistic Regression model trained on a public data source\,\footnote{\,\url{https://www.kaggle.com/datasets/crowdflower/twitter-user-gender-classification}}.
    \item Readability features: a Python implementation of several readability measures\,\footnote{\,\url{https://github.com/andreasvc/readability/}}.
    \item Sentiment features: VADER module from NLTK\,\footnote{\,\url{https://www.nltk.org/index.html}}.
    \item Emotion \& Hate-Speech features: A Transformer-based library for SocialNLP tasks\,\footnote{\,\url{https://github.com/pysentimiento/pysentimiento}}.
\end{itemize}

\begin{table}[!t]
\centering
\small
\renewcommand{\arraystretch}{1.05}
\caption{Message (M) Features.}
\begin{tabular}{l|ll}
\hline\hline
\textbf{Sub-type} & \textbf{Feature Name} & \textbf{Description} \\ 
(\# of features) &   &  \\
\hline
\multirow{13}{*}{\textbf{Topic} (39)} & M\_TopicM1 & Likelihood that the post belongs to a specific topic in the 19-topic scheme\,$^1$.\\ 
& ... & ...\\ 
&M\_TopicM19 & (Same as above)\\ 
&M\_TopicMMain  & The most likely topic with the highest likelihood in the 19-topic scheme. \\ 
&M\_TopicMNum & The number of identified topics with likelihood \ensuremath{>} 0.5 in the 19-topic scheme. \\ 
&M\_TopicG1 & Likelihood that the post belongs to a specific topic in the 6-topic scheme\,$^2$. \\ 
& ... & ... \\ 
&M\_TopicG6 & (Same as above) \\ 
&M\_TopicGMain  & The most likely topic with the highest likelihood in the 6-topic scheme. \\ 
&M\_TopicGNum & The number of identified topics with likelihood \ensuremath{>} 0.5 in the 6-topic scheme. \\ 
&M\_TopicLDA1 & Likelihood that the post belongs to 1 of the 10 topics\,$^3$ identified by the LDA model. \\ 
& ...&...\\ 
&M\_TopicLDA10 & (Same as above) \\ 
\hline
\multirow{10}{*}{\textbf{Language} (10)} & M\_CharNum & The number of characters of the post. \\ 
&M\_WordNum & The number of words of the post.\\ 
&M\_Grammar1 & Scores for grammatical and spelling correctness on the level of words. \\ 
&M\_Grammar2 & Scores for grammatical and spelling correctness of the whole post. \\ 
&M\_Polarity & The orientation of the expressed tone of the post. \\ 
&M\_Subjectivity & A measure of personal opinion and factual information contained in the post. \\ 
&M\_Irony & A measure of how ironic the post is. \\ 
&M\_Offensive & A measure of how offensive the post is. \\ 
&M\_Emoji & The most likely emoji that describes the textual content. \\ 
&M\_Masculinity & Whether the language style of the text is masculine. \\ 
\hline
\multirow{5}{*}{\textbf{Readability} (11)} & M\_Readability1 & Different measures of how difficult the post in English is to understand. \\ 
&... & (Readability measures include: Kincaid, ARI, Coleman-Liau, FleschReadingEase, \\ 
&M\_Readability9 &  \,\,\,\,\,\,GunningFogIndex, SMOGIndex, LIX, RIX, DaleChallIndex) \\ 
&M\_Readability10 & Count of words in the post that consist of three or more syllables. \\ 
&M\_Readability11 & Count of words in the post that are not on the list used in DaleChallIndex. \\ 
\hline
\multirow{5}{*}{\textbf{Sentiment} (5)} & M\_Sentiment1 & Negative sentiment score of the post. \\ 
&M\_Sentiment2 & Neutral sentiment score of the post. \\ 
&M\_Sentiment3 & Positive sentiment score of the post. \\ 
&M\_Sentiment4 & Compound value of M\_sentiment1, M\_sentiment2, and M\_sentiment3. \\ 
&M\_SentimentMain & The most likely sentiment label according to M\_sentiment4.\\ 
\hline
\multirow{4}{*}{\textbf{Emotion} (8)} & M\_Emotion1 & The probability that the post expresses a specific type of emotion. \\ 
&... & (Emotions studied: anger, joy, fear, disgust, surprise, sad, others.) \\ 
&M\_Emotion7 & (Same as above) \\ 
&M\_EmotionMain & The most likely emotion expressed by the post. \\ 
\hline
\multirow{4}{*}{\textbf{Hate-Speech} (4)} & M\_Hate1 & A measure of aggressiveness of the post. \\ 
&M\_Hate2 & A measure of hatefulness of the post. \\ 
&M\_Hate3 & A measure of whether the post is targeting an individual (or entity). \\ 
&M\_HsNum & The \# of the above types of hate speech (above given thresholds) detected in the post.  \\ 
\hline
\textbf{Hashtag} (1) &M\_hashtag & The categorical label of the hashtag under study contained in the post. \\ 
\hline\hline
\multicolumn{3}{p{0.95\linewidth}}{\footnotesize 
$^1$\,The 19 topic-scheme includes: arts\&culture, business\&entrepreneurs, celebrity\&pop\_culture, diaries\&daily\_life, family, fashion\&style, film\_tv\&video, fitness\&health, food\&dining, gaming, learning\&educational, music, news\&social\_concern, other\_hobbies, relationships, science\&technology, sports, travel\&adventure, youth\&student\_life.
} \\
\multicolumn{3}{p{0.95\linewidth}}{\footnotesize 
$^2$\,The 6 topic-scheme includes: arts\&culture, business\&entrepreneurs, pop\_culture, daily\_life, sports\&gaming, science\&technology. 
} \\
\multicolumn{3}{p{0.95\linewidth}}{\footnotesize 
$^3$\,The top 3 keywords for each topic discovered by the LDA model: 1) blm, amp, last; 2) nhs, covid, go; 3) john, mp, dido; 4) nhs, ios, one; 5) new, china, america; 6) loveisland, davide, think; 7) energy, brexit, energyprices; 8) avengers, marvel, news; 9) monkeypox, ukraine, russian; 10) covid, youtube, supercup. 
}
\end{tabular}
\label{app:Mfeatures}
\end{table}

\begin{table}[!t]
\centering
\small
\renewcommand{\arraystretch}{1.05}
\caption{User Profile (U-P) Features}
\begin{tabular}{c|ll}
\hline\hline
\multicolumn{3}{l}{\multirow{2}{*}{\bf For Recipient (R)}} \\
\multicolumn{3}{l}{}\\
\hline 
\textbf{Sub-type} & \textbf{Feature Name} & \textbf{Description} \\ 
(\# of features) &   &  \\ 
\hline
& U-P\_R\_AccountAge & The number of days since R's account was registered (until 01/09/2022). \\ 
& U-P\_R\_FollowerNum & The number of R's followers. \\ 
& U-P\_R\_FolloweeNum & The number of R's followees (users that R is following). \\ 
& U-P\_R\_TweetNum & The number of total posts that R has posted since registration. \\ 
& U-P\_R\_ListedNum & The number of lists R is included in. \\ 
& U-P\_R\_SpreadActivity & The ratio of the \# of R's posts to the maximal \# of posts among all users. \\ 
\textbf{Profile} (12) & U-P\_R\_FollowerNumDay & The number of followers normalised by the age of this account. \\ 
& U-P\_R\_FolloweeNumDay & The number of followees normalised by the age of R. \\ 
& U-P\_R\_TweetNumDay & The number of total posts normalised by the age of this account. \\ 
& U-P\_R\_ListedNumDay & The number of lists that the user is included in normalised by the account age. \\ 
& U-P\_R\_ProfileVerified & A binary label (0 or 1) indicating R is verified by X (1) or not (0). \\ 
& U-P\_R\_ProfileUrl & A binary label (0 or 1) indicating a URL is shown in R's profile (1) or not (0). \\ 
\hline
& U-P\_R\_LeaderRank & A value to measure the influence of R. A variation of PageRank. \\ 
\textbf{Network} (3) & U-P\_R\_Indegree & The number of users who have followed and interacted with R. \\ 
& U-P\_R\_FollowS & A binary label (0 or 1) indicating R is following S (1) or not (0). \\ 
\hline
\multicolumn{3}{l}{\multirow{2}{*}{\bf For Sender (S) }} \\
\multicolumn{3}{l}{}\\
\hline 
\multicolumn{3}{c}{(Same as above except replacing `\_R\_' with `\_S\_' in feature names and replacing `R' with `S' in description.)} \\ 
\hline\hline
\end{tabular}
\label{app:Ufeatures}
\end{table}

\begin{table}
\centering
\small
\renewcommand{\arraystretch}{1.05}
\caption{User Historical Action (U-HA) Features}
\begin{tabular}{c|ll}
\hline\hline
\multicolumn{3}{l}{\multirow{2}{*}{\bf For Recipient (R)}} \\
\multicolumn{3}{l}{}\\
\hline
\textbf{Sub-type} & \textbf{Feature Name} & \textbf{Description} \\
(\# of features) &   &  \\ 
\hline
& U-HA\_R\_TweetNum & The number of R's historical posts. \\ 
& U-HA\_R\_TweetPercent & Percentage of original-tweets/retweets/quotes/replies in R's historical posts.\\ 
& U-HA\_R\_RetweetPercent & (See above)\\ 
\textbf{Activity} (7) & U-HA\_R\_QuotePercent & (See above)\\ 
& U-HA\_R\_ReplyPercent & (See above)\\ 
& U-HA\_R\_InteractivePer & Percentage of reposts in R's historical posts. \\ 
& U-HA\_R\_AverageInterval & The average time interval (days) between the posting of R's historical posts. \\ 
\hline
& U-HA\_R\_RetweetedRate & Average number of retweets/quotes/replies/likes received by R's historical posts. \\ 
\textbf{Popularity} (4) & U-HA\_R\_QuotedRate & (See above)\\ 
& U-HA\_R\_RepliedRate & (See above)\\ 
& U-HA\_R\_LikedRate & (See above)\\ 
\hline
\multicolumn{3}{l}{\multirow{2}{*}{\bf For Sender (S)} } \\
\multicolumn{3}{l}{}\\
\hline
\multicolumn{3}{c}{(Same as above except replacing `\_R\_' with `\_S\_' in feature names and replacing `R' with `S' in description.)} \\ 
\hline
\multicolumn{3}{l}{\multirow{2}{*}{\bf For both R and S} } \\
\multicolumn{3}{l}{} \\
\hline
& U-HA\_RS\_Mention & The number of times that R mentioned S in R's historical posts.\\ 
& U-HA\_RS\_MentionPer & Percentage of posts in which R mentioned S in R's historical posts.\\ 
& U-HA\_SR\_Mention & The number of times that S mentioned R in S's historical posts.\\ 
& U-HA\_SR\_MentionPer & Percentage of posts in which S mentioned R in S's historical posts.\\ 
\textbf{Interaction}& U-HA\_RS\_RepostLatency & Latency (days) from the creation time of the post to the reposting. \\ 
(16)& U-HA\_SR\_TORS1 & A multi-dimensional vector to depict users’ relationship under different topics. \\ 
& ... & ... \\ 
& U-HA\_SR\_TORS10 & (Same as above) \\ 
& U-HA\_SR\_PathWidth & The cosine distance between the post’s topic vector and the Topic-Oriented \\ 
& & Relationship Strength between S and R. \\ 
\hline\hline
\end{tabular}
\label{app:HUfeatures}
\end{table}

\end{document}